\documentclass[useAMS,usenatbib]{mn2e}
\usepackage{graphicx}

\def\msun{M_\odot}
\def\fs{f_{\rm S}}
\def\fsi{f_{\rm S_I}}
\def\I0{I_{\rm 0}}
\def\tE{t_{\rm E}}
\def\te{t_{\rm E}}
\def\t0{t_{\rm 0}}
\def\u0{u_{\rm 0}}

\newcommand{\eg}{{e.g.},\,}
\newcommand{\ie}{{i.e.},\,}
\newcommand{\apj}{{Astrophysical Journal}}
\newcommand{\apjl}{{Astrophysical Journal Letters}}

\newcommand{\araa}{{ARA\&A}}
\newcommand{\aap}{{Astronomy \& Astrophysics}}
\newcommand{\mnras}{{MNRAS}}
\newcommand{\aj}{{\apj}}

\title[The OGLE-II View of Microlensing towards the SMC.]
{The OGLE View of Microlensing towards the Magellanic Clouds. II. OGLE-II SMC data.  \thanks{Based on
    observations obtained with the 1.3~m Warsaw telescope at the Las Campanas Observatory of the Carnegie Institution of Washington.}}
\author[{\L}. Wyrzykowski et al.]
{{\L}. Wyrzykowski$^{1,2}$\thanks{email: wyrzykow@ast.cam.ac.uk, name
    pronunciation: {\it Woocash Vizhikovsky}}, S. Koz{\l}owski$^3$,
  J. Skowron$^2$, V. Belokurov$^1$, M. C. Smith$^1$, \newauthor
  A. Udalski$^2$, M. K. Szyma{\'n}ski$^2$, M. Kubiak$^2$,
  G. Pietrzy{\'n}ski$^{2,4}$,
  \newauthor I. Soszy{\'n}ski$^2$, O. Szewczyk$^{2,4}$\\
  $^1$ Institute of Astronomy, University of Cambridge,
  Madingley~Road,
  Cambridge~CB3~0HA,~UK \\
  $^2$ Warsaw University Astronomical Observatory, Al.~Ujazdowskie~4, 00-478~Warszawa, Poland \\
  $^3$ Department of Astronomy, Ohio State University, 140 W. 18th Ave., Columbus, OH 43210, USA\\
  $^4$ Universidad de Concepci{\'o}n, Departamento de Fisica, Astronomy Group, Casilla 160-C, Concepci{\'o}n, Chile\\ }

\begin{document}

\date{Accepted  2010 April 28. Received 2010 April 19; in original form 2009 November 10  }

\pagerange{\pageref{firstpage}--\pageref{lastpage}} \pubyear{2010}

\maketitle

\label{firstpage}

\begin{abstract}

The primary goal of this paper is to provide the evidence that can either prove or falsify the hypothesis that dark matter in the Galactic halo can clump into stellar-mass compact objects. If such objects existed, they would act as lenses to external sources in the Magellanic Clouds, giving rise to an observable effect of microlensing. 
We present the results of our search for such events, based on the data from the second phase of the OGLE survey (1996-2000) towards the SMC.
The data set we used is comprised of 2.1 million monitored sources distributed over an area of 2.4 square 
degrees. 
We found only one microlensing event candidate, however its poor quality light curve limited our discussion on the exact distance to the lensing object. 

Given a single event, taking the blending (crowding of stars) into account for the detection efficiency simulations, 
and deriving the {\it HST}-corrected number of monitored stars, the microlensing optical depth is $\tau=(1.55\pm1.55) \times 10^{-7}$. This result is 
consistent with the expected SMC self-lensing signal, with no need of introducing dark matter microlenses.
Rejecting the unconvincing event leads to the upper limit on the fraction of dark matter in the form of MACHOs to $f<20$ per cent for deflectors' masses around 0.4 $\msun$ and $f<11$ per cent for masses between 0.003 and 0.2 $\msun$ (95 per cent confidence limit).
Our result indicates that the Milky Way's dark matter is unlikely to be clumpy and form compact objects in the sub-solar-mass range. 

%%%
\end{abstract}

\section{Introduction}

The Magellanic Clouds are harbours to millions of stars. The light of each of these objects can be magnified if another massive 
object is close enough to the line-of-sight connecting the observer and a distant star. 
\citet{Paczynski1986} first realised that with the advent of CCDs, forthcoming massive photometric surveys could 
effectively test the hypothesis that dark matter in the Galactic halo can clump and form Massive Compact Halo Objects (MACHOs). 
These objects, if they existed, would act as lenses to more distant LMC/SMC stars, within the reach of current observing facilities. 
This brilliant, yet simple idea triggered several microlensing programs to emerge. The first detections of the microlensing effect were reported by the  
MACHO \citep{MACHO}, OGLE \citep{Udalski1993}, EROS \citep{EROS}, MOA \citep{MOA}, 
Angstrom \citep{ANGSTROM}, POINT-AGAPE \citep{POINTAGAPE}, 
and WeCaPP \citep{WECAPP} microlensing teams. 

For almost two decades, microlensing as an astrophysical tool has been very successful in finding objects which do not emit any or emit little light.
The OGLE group alone have discovered over 4000 ordinary microlensing events to date. A list of exotic microlensing events includes detection of
black-holes 
(\eg \citealt{OGLEBH}), 
planets (\eg \citealt{UdalskiOB05071}, \citealt{Gaudi2008}, \citealt{DongKB07400}), 
binary stars \citep{Skowron2007binaries} and also a variety of effects such as
the parallax (\eg \citealt{Smith2003parallax}, \citealt{Gould2009terrestialParallax}), 
xallarap \citep{Assef2006MACHO97SMC1}, etc.

However, since the microlensing field has evolved into a tool nowadays primarily concentrated on finding either the most distant or the smallest known planets,
Paczy{\'n}ski's original idea has been somewhat forgotten. The primary motivation for this paper is to fully explore the
existing OGLE data to search for microlensing events towards the Magellanic Clouds. As of 2010 we have collected approximately 13 seasons (4 seasons of OGLE-II
and 9 seasons of OGLE-III) of data for both the LMC and SMC. In \cite{Wyrzykowski2009} (hereafter Paper I) we presented our first estimate 
of the microlensing optical depth towards the LMC from the OGLE-II data. The detection of two events led to the optical depth of 
$\tau_{\rm LMC} = (0.43\pm0.33)\times 10^{-7}$. 
However, the MACHO collaboration derived the optical depth of $\tau_{\rm LMC} = (1.0 \pm
0.3) \times 10^{-7}$ based on their 10 candidates (\citealt{AlcockMACHOLMC}, \citealt{BennettMACHOLMC}). 
If this number is compared to the optical depth for the Galactic halo entirely made of MACHOs, $\tau_{\rm halo} \approx 4.7 \times 10^{-7}$ (\citealt{BennettMACHOLMC}), 
it gives the fractional contribution of $f = \tau_{\rm LMC}$/$\tau_{\rm halo} \approx$ 20 per cent.
On the other hand, the EROS collaboration has derived $\tau_{\rm LMC} < 0.36 \times 10^{-7}$, which translates to $f<8$ per cent only \citep{TisserandEROSLMC}.
The OGLE--II estimate of $f<10$ per cent from Paper I, favours the EROS solution but the two detected events are 
also consistent with the expected LMC self-lensing signal. 

The SMC has received somewhat less attention in terms of microlensing studies than the LMC.
So far, only the EROS data were studied systematically and the optical depth of $\tau_{\rm SMC} =  (1.7 \pm 1.7) \times10^{-7}$ was derived for one microlensing event detected in their Bright Stars Sample \citep{TisserandEROSLMC}. 
Another study of 5 years of the EROS data gives $f<25\%$ for objects with masses $10^{-7}$ to $1\msun$ \citep{EROSSMC2003}. 
On the other hand, the MACHO collaboration estimated the optical depth to be $(2-3) \times 10^{-7}$ based on their two events \citep{Alcock1999MACHO98SMC1}.
The SMC self-lensing estimates are in a range of $(0.4$--$1.8) \times10^{-7}$ from N-body simulations by \cite{Graff1999} and from analytical work of \cite{EROSSMC1998}.

In this paper we extend our work from Paper I, on search for dark matter compact objects in the Galactic halo, to an 
independent SMC data set collected by OGLE during its second phase in years 1996--2000.
The paper has the following structure.  First, the empirical optical depth estimator is described. 
Then the observational data used in the analysis are presented in Section \ref{sec:data}.
Next, in Section \ref{sec:search} the search procedure for events is described and its yield presented. 
The detection efficiency of events and the calculation of the optical depth is discussed in Section 
\ref{sec:results}--\ref{sec:tau}. The paper concludes with a discussion of the results.

\section{Experimental Optical Depth}

A review of the microlensing-related quantities is given in \cite{Paczynski1986,Paczynski1996} and 
\citep{Gould2000microlensingFormalism}.
In short, the time-scale (Einstein radius crossing time) $\tE$ is the only physical parameter of an event that, in the simplest case, 
can be derived while fitting an observed light curve with the microlensing model. The basic point-lens point-source 
microlensing light curve \citep{Paczynski1996} is described by

\begin{equation}
\label{eq:I}
\textrm{mag(t)} = \textrm{mag}_0 - 2.5\log\left[ \fs A(t) + (1-\fs)\right],
\end{equation}
where $\textrm{mag(t)}$ is the observed magnitude at a given moment of time $t$, $\textrm{mag}_0$ is the baseline magnitude (away from the peak) and
$\fs$ is the ratio of the lensed source flux to the total flux of stars within the seeing disk (blending parameter).
The amplification $A(t)$ then depends on the lensing geometry, which changes with time, and can be evaluated with the equations
\begin{equation}
\label{eq:A}
A(t) = { u(t)^2 + 2 \over u(t)\sqrt{u(t)^2+4} } \quad {\rm and} \quad u(t) = \sqrt{\u0^2 + {{(t-\t0)^2} \over {\tE^2}}},
\end{equation}
where $\u0$ is the impact parameter and $\t0$ is the time of the maximum of the peak.

We calculate the empirical optical depth following \citep{Paczynski1996} as

\begin{equation}
\label{eq:tau}
\tau={{\pi}\over {2 N_* T_{\rm obs}}}\displaystyle \sum_i^{N_{\rm ev}} {{\tE}_{i} \over \epsilon({\tE}_i)},
\end{equation}

\noindent where $T_{\rm obs}$ is the time-span of all observations,
$N_*$ is the total number of monitored stars, 
$N_{\rm ev}$ is the total number of events, 
${\tE}_i$ is the time-scale of individual events detected with the efficiency of
$\epsilon({\tE}_i)$.  
The non-trivial parts of the estimator are the total number of monitored stars and the detection efficiency, 
and they are described in detail in Section \ref{sec:blending}.

\begin{figure*}
\includegraphics[width=12cm]{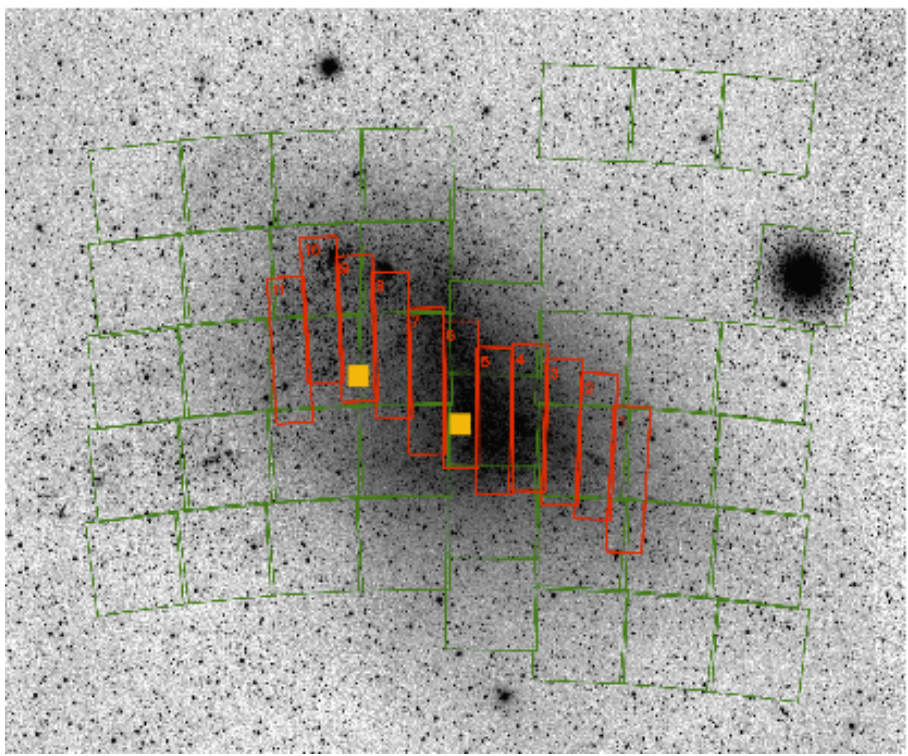}
\caption{Positions of the OGLE--II SMC fields (red rectangles with
  labels). Also shown are all OGLE--III fields (green squares).
  The two small filled squares show  the positions of the {\it HST} fields used for our blending determination.
  Background image credit: ASAS all sky survey.}
\label{fig:fields}
 \end{figure*}

\begin{table*}
\centering
\caption{Summary of the OGLE--II SMC fields.}
\label{tab:fields}
\begin{tabular}{cccrrcc}
\hline
\noalign{\vskip5pt}
Field & $RA_{J2000}$ & $Dec_{J2000}$ & \multicolumn{2}{c}{No of stars} & Stellar density & Density level\\
& & & template & estimated real & [stars/sq.arcmin] & \\
\noalign{\vskip5pt}
\hline
\noalign{\vskip5pt}
SMC\_SC1 & 0:37:50.89 & -73:29:42.0 & 96,220 & 158,894 & 123 & sparse \\
SMC\_SC2 & 0:40:53.11 & -73:17:29.0 & 131,911 & 217,668 & 168 & sparse \\
SMC\_SC3 & 0:43:57.88 & -73:12:29.0 & 196,073 & 324,564 & 250 & sparse \\
SMC\_SC4 & 0:46:58.59 & -73:07:29.8 & 246,137 & 406,992 & 314 & sparse \\
SMC\_SC5 & 0:50:00.59 & -73:08:46.0 & 306,784 & 591,366 & 391 & dense \\
SMC\_SC6 & 0:53:00.77 & -72:58:40.7 & 303,571 & 585,153 & 387 & dense \\
SMC\_SC7 & 0:55:59.64 & -72:53:32.5 & 235,195 & 389,400 & 300 & sparse \\
SMC\_SC8 & 0:58:57.59 & -72:39:30.7 & 188,159 & 311,264 & 240 & sparse \\
SMC\_SC9 & 1:01:54.61 & -72:32:32.5 & 159,055 & 263,061 & 203 & sparse \\
SMC\_SC10 & 1:04:50.47 & -72:24:47.2 & 132,781 & 219,236 & 169 & sparse \\
SMC\_SC11 & 1:07:45.40 & -72:39:32.3 & 110,230 & 182,010 & 141 & sparse \\
\hline
total & \multicolumn{2}{c}{} & 2,106,303 & 3,649,608 & \\
\noalign{\vskip5pt}
\hline
\end{tabular}

\medskip
\begin{flushleft}
{\it Note:} Coordinates point to the centre of the field, each being $14' \times
56'$. Number of ``good'' objects in the template is provided ($N>80$ and
$\langle I \rangle < 21.0$ mag) together with the estimated number of
real monitored stars (see Section \ref{sec:blending}). Stellar density
in number of stars per square arc minute was used to classify fields
into {\it dense} or {\it sparse} classes with the threshold of 300 stars/sq.arcmin.
\end{flushleft}
\medskip
\bigskip
\bigskip
\end{table*}

\section{Observational Data}
\label{sec:data}

\begin{table}
\centering
\caption{Error correction coefficients for OGLE--II SMC fields.}
\label{tab:errorcor}
\begin{tabular}{ccc}
\hline
Field & $\gamma$ & $\epsilon$ \\
\hline
SMC\_SC1 & 1.195372 & 0.003250 \\  
SMC\_SC2 & 1.223286 & 0.001327 \\  
SMC\_SC3 & 1.277762 & 0.002759 \\  
SMC\_SC4 & 1.153056 & 0.002099 \\  
SMC\_SC5 & 1.218417 & 0.002249 \\  
SMC\_SC6 & 1.214665 & 0.001926 \\  
SMC\_SC7 & 1.233638 & 0.002660 \\  
SMC\_SC8 & 1.190629 & 0.002621 \\  
SMC\_SC9 & 1.182015 & 0.002673 \\  
SMC\_SC10 & 1.167742 & 0.002901 \\  
SMC\_SC11 & 1.222095 & 0.002335 \\  
\hline
\end{tabular}
\end{table}

\begin{table*}
\centering
\caption{Selection criteria for search for microlensing events in the OGLE--II data}
\label{tab:conditions}
\begin{tabular}[h]{c|lrr}
\hline
Cut no. & & & No. of objects left \\
\hline     
0 & Selection of ``good'' objects & $N>80$, $\langle I \rangle \le 21.0$ mag  &  2,106,303\\
& & \\

1 & Significant bump over baseline & $\displaystyle\sum_{peak} \sigma_i > 30.0 $ & 3,097 \\
& & \\

2 & ``Bumper'' cut$^\dagger$ & $\langle I \rangle>19.0~mag $,  $\langle V-I \rangle>0.5~mag$ & 2,295 \\ 
& &  &\\

3 & Microlensing fit better than constant line fit & ${{\chi^2_{line}-\chi^2_{{\mu} 4}}\over{{\chi^2_{{\mu}4}\over{N_{dof,\mu4}}}\sqrt{2N_{dof,\mu4,peak}}}} > 140$  & 335 \\
& & & \\

4 & Number of points at the peak$^{\ddag}$ & $N_{peak} > 6$  & 311 \\ 
& & & \\

5 & Microlensing fit better than supernova fit & $\chi^2_{SN} > min(\chi^2,\chi^2_{{\mu}4})$ & 201 \\
& & & \\

6 & Peak within the data span   & $466 \le {\t0} \le 1874$   & 192 \\
& [HJD-2450000]    &         &         \\
& &  & \\

7 & Blended fit converged & $0< \fs < 1.2$ & 44 \\
& & & \\

8 & Conditions on goodness of microlensing fit &  ${{\chi^2}\over{N_{dof}}} \le 2.3 $ and ${{\chi_{\mu4,peak}^2}\over{N_{dof,\mu4,peak}}} \le 5 $ & 2(1)$^\ast$ \\
& (global and at the peak) & & \\

9 & Time-scale cut & $1 \le {\tE} \le 500$ & 2(1)$^\ast$ \\
& $[d]$ &  & \\
&  & & \\

10 & Impact parameter cut &  $0 < {\u0} \le 1 $ & 2(1)$^\ast$ \\
& & & \\

\hline
\end{tabular}
\\
\begin{flushleft}
$^{\dagger}$ magnitudes as in the field SMC\_SC1 (shifted according to the position of the centre of Red Clump) \\
$^{\ddag}$in the range of $\t0 \pm 1 \tE$ \\
$^{\ast}$ there was in fact only one event occurring, whose flux was detected on two template objects (see text)
%$\chi_{\mu}$ is non-blended fit \\
%$\chi_{{\mu}5}$ is blended fit \\
\end{flushleft}
\end{table*}

\begin{table*}
\caption{OGLE--II database stars in the SMC on which the flux from the microlensing candidate was detected.}
\label{tab:objects}
\begin{center}
\begin{tabular}{ccccccc}
\hline
field & DB             &    \multicolumn{2}{c}{image} & \multicolumn{2}{c}{baseline}  \\
         & star id                   &          x [pix] & y [pix]      & $I$ [mag]                 & $V$ [mag]               \\
\hline
SMC\_SC7 & 193726 &  1517.87   & 6448.29 & 15.34$\pm$0.01 & 16.48$\pm$0.02 \\
& & & & &  \\
	
SMC\_SC7 & 351179 &  1515.84 &  6448.42 & 19.7$\pm$0.2 & 21.9$\pm$0.5 \\  	

\hline			       
\end{tabular}
\end{center}
\end{table*}

\begin{figure}
\includegraphics[width=8cm]{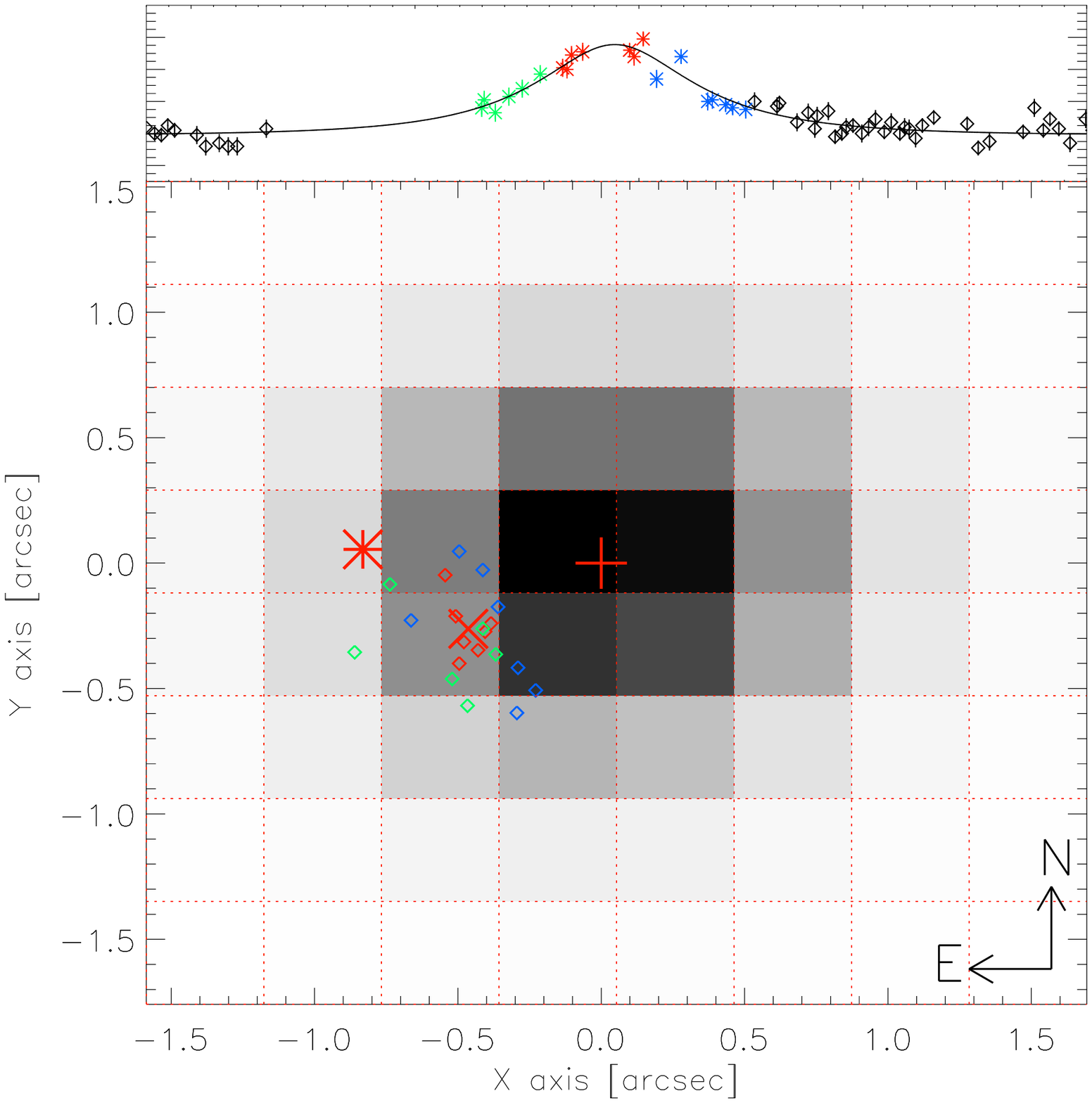}
\caption{Astrometric analysis of the OGLE-SMC-01 microlensing event is shown. {\it Top panel} shows the light curve of the event, divided into parts: rising (green), maximum (red) and falling (blue). 
{\it Bottom panel}  shows the $8\times8$ pixel area from the OGLE-II template, centred close to the event position. The range of grey levels starts from the lack of flux (white) and increase towards dark levels with the increasing flux. 
The centroids of the bright and faint stars from the OGLE-II catalogue, on which the event was detected, are marked with the plus and star, respectively. 
With the colour coding from the top panel, we show the positions of the magnified light measured on the subtracted images. The
$\times$ symbol shows the mean position derived from the individual positions. This event location is inconsistent with any of the two detected stars, meaning that there must be yet another, third object that was microlensed.
}
\label{fig:astrometry}
\end{figure}

The data used in this work were gathered during the second phase of the OGLE survey, which started in 1996 and lasted until 2000. The project used its own dedicated Warsaw Telescope located in the Las Campanas Observatory, Chile. The details on the instrumentation for the OGLE--II survey can be found in \citet{OGLE2}.

A single 2k$\times$2k CCD chip (pixel size of 0.417~arcsec) was
operated in the drift-scan mode, giving an actual size for each observed field
of 2k$\times$8k (14$\times$56~arcmin).
There were 11 fields observed towards the SMC covering in total 2.4 square degrees. Their locations are shown in
Fig. \ref{fig:fields} marked over an image of the SMC obtained by the ASAS survey\citep{ASAS}.  Table \ref{tab:fields} lists all the fields with the coordinates of their centres, number of good objects in
$I-$band, blending-corrected number of stars (see Section \ref{sec:blending}) and blending density group. 
By ``good'' we mean all objects having at least 80 observations during the entire time span of the OGLE-II (from about 22 to 27 per cent of all collected frames of a field) and mean magnitude brighter than $21.0$ mag. The limiting magnitude
was chosen at the peak of the observed luminosity functions.
Similarly as in Paper I, fields were divided into dense and sparse according the density of objects.
Here we set the density boundary at 300 ``good'' stars per sq. arcmin, slightly less than in Paper I, in order to diversify the density levels in the SMC.
The density levels correspond to respective {\it HST} images which were used for deriving the blending distributions (see Section \ref{sec:blending}).

The OGLE--II observations of the SMC began in January 1997
(HJD=2450466) with the monitoring of the SC5 and SC6 fields only. 
Nearly half a year later (HJD=2450621) the remaining nine fields were added to the observing queue.
All eleven fields were monitored continuously until November 2000 (HJD=2451874) 
yielding about 300 frames per field in $I$-band.  
Additionally, there were 30 to 45 frames per field collected in $V$-band. 
On average each field was observed every third night in $I$-band and
every 8-th night in $V$-band, with the mean seeing of 1.36 arcsec in $I$ and
1.39 arcsec in $V$. Raw images were de-biased and flat-field
corrected ``on-the-fly''. The photometric pipeline was based on the
Difference Image Analysis method (DIA, \citealt{WozniakDIA, AlardLuptonDIA}). 
The template images for DIA were created by stacking the best
quality images, resulting in images with the seeing of about 1.1 arcsec.
Photometric databases were designed for both
pass-bands as described in \citet{Szymanski2005} and are available on-line\footnote{http://ogledb.astrouw.edu.pl}.
The photometry of the template images contain about 2.1 million objects in $I-$ and $V-$band colours suitable for our study.
The search for microlensing events and determination of the detection efficiency 
were performed using the $I-$ band data only, as these
were far more numerous and were sampled more frequently as compared to the $V-$band light curves. 
In this paper we also occasionally use additional data collected during the third phase of OGLE (OGLE--III), years 2001--2009. 
The detailed description of the OGLE-III instrumentation and photometric pipeline can be found in \citet{EWSOGLE3}.
The search for microlensing events in the OGLE--III SMC data will be presented separately in a 
forthcoming paper (Wyrzykowski et al. in preparation).

Before we start the search procedure, all $I$-band photometric measurements have to have their error-bars 
corrected, since DIA is known to underestimate the photometric error-bars (\citealt{WozniakDIA}). The correction technique is
detailed in Paper I. In short, the method compares the intrinsic error-weighted {\it rms} of constant stars in each 
field with their mean error. The error-bars are corrected with the formula
\begin{equation}
\Delta I_{cor} = \sqrt{ ( \gamma \Delta I )^2 + \epsilon^2 },
\label{eq:errors}
\end{equation}
where $\Delta I_{cor}$ is the corrected error-bar, $\Delta I$ is the original error-bar returned by the photometry pipeline, and
$\gamma$ and $\epsilon$ are the correction coefficients.
They are derived for each SMC field and are presented in Table \ref{tab:errorcor}.
The mean $\gamma$ and $\epsilon$ for all SMC fields were $1.20715$ and $0.002436$, respectively.

%%%%%%%%%%%%%%% SEARCH PROCEDURE %%%%%%%%%%%%%
\section{Search procedure}
\label{sec:search}

The main objective of this paper is to find rare microlensing events amongst millions of SMC stars. The task requires a number of steps to 
iteratively remove all unwanted, contaminating light curves.

Our search procedure begins by pulling out light curves from the database with more than 80 epochs and the mean $I$-band magnitude 
brighter than 21.0 mag (Cut 0). This sample of 2.1 million light curves becomes our All Stars Sample. 
The following steps along with the number of objects left after each cut are detailed in Table \ref{tab:conditions}.
We use the same search pipeline as in Paper I, however, the exact parameters of the cuts were fine-tuned independently 
for OGLE--II SMC data with the Monte Carlo simulations.
One of the main differences is in Cut 0, where the depth of the search has changed from $20.4$ mag (LMC) to $21.0$ mag (SMC) due 
to a lessen crowding and greater distance modulus. 
Another important change, as compared to Paper I, is the definition of the ``blue bumper'' region (Cut 2). The magnitude limit is chosen to be $0.5$ mag fainter, than for the LMC, due to larger distance modulus of the SMC. 

A potential source of contamination in a sample of microlensing events can be blue bumpers and supernovae. Both these classes are characterised by
similar light curves to that of microlensing, especially if a light curve is of low photometric quality or sparsely sampled. 
Blue bumpers can be effectively removed by using Cut 2, as they occupy the bright blue end of the main sequence.
Also blue bumpers are known to exhibit several bumps on a time-scale of a few years. This is why we make use of the OGLE-III data, giving us a full span of $\sim13$ years,
to understand the nature of each microlensing event candidate.
The contamination with supernovae in the OGLE--II is very limited due to the fact that we observed a very small central region of the SMC.
There should be statistically 1 (4) SNe peaking above $I<20$ (21) mag in our data.
In Cut 5, we compare the goodness of fit of the microlensing and supernova models, removing $\sim100$ light curves with an asymmetric bump.

% In the years 1998-2000 the OGLE-II SMC data were analysed in real-time with the OGLE's Early 
% Warning System \citep{Udalski1994ews}, however, the system did not detect any candidate microlensing eve

%%%%%%%%%%%%%%%%%%%%%%%%% RESULTS %%%%%%%%%%%%%%%%
\section{Search Results}
\label{sec:results}

Our search pipeline has returned two microlensing event candidates. The close inspection of their positions, 
however, revealed they are separated from each other by 2 pixels ($\sim 0.9$ arcsec) only. 
Moreover, the moment of the maximum brightness was exactly the same in both light curves. 
Table \ref{tab:objects} presents detailed information regarding both objects, including their positions and magnitudes.
The probability for two independent microlensing events occurring at the same moment of time in such proximity is close to zero.
We conclude, therefore, that both these events are the image of the same single microlensing event into two nearby stars.

To pinpoint the unbiased position of the event we employed DIA.
Since the template image is matched astrometrically and photometrically to each image in a series, what is left on subtracted images are
only variable objects (with either positive or negative fluxes) and systematic problems. All constant objects are subtracted out.
Fig. \ref{fig:astrometry} shows astrometric measurements of the centroid of the differential flux at various stages 
of the event (indicated with different colours). 
It is evident that the most amplified data points from the peak of the event (also with the most accurate astrometry) 
are concentrated away from the centroids of both ``ghost'' microlensing events (marked with a plus and a star).
This indicates that the microlensed flux is not linked with neither of these objects and there must be yet another 
star hidden in the wings of the brighter object. This object is not present on the template.
We repeated the DIA analysis with a prior knowledge of the exact source location on the CCD chip ($x,y=1516.74, 6447.66$) 
and measured the flux at that position. This microlensing event was dubbed OGLE-SMC-01.

%%%%%%%%%%%%%%%%%%%%%%%%%%%%%%%%%%%%%%%%%%%%%%%%%% table
\begin{table*}
\caption{ Parameters of the microlensing model fits to the event OGLE-SMC-01. Four- and five-parameter fits were performed on the OGLE--II $I$-band data solely. 
The seven-parameter fit used also the $V$-band measurements (obtained from transformed MACHO-$B$ data)  and included the baseline variability model with period of 1 yr.}
\label{tab:fits}
%{\scriptsize
\begin{tabular}[h]{lrlrlrl}
\hline
\multicolumn{7}{c}{OGLE-SMC-01}\\
parameter & \multicolumn{2}{c}{four-parameter fit} & \multicolumn{2}{c}{five-parameter fit} & \multicolumn{2}{c}{seven-parameter fit}\\
\hline
%parameter	& value	& $\sigma$	&  value & $\sigma$ & value & $\sigma$ \\
%\hline
$\t0$	\dotfill	& 1385.7  & $\pm1.4$	& 		1385.9	& $\pm1.4$ & 1384.10 & $\pm0.78$ \\
 & & & & & &\\
$\tE$	\dotfill	& 31.5	& $\pm1.2$		&  65.0	& $\pm21.8$ & 89.7 & $\pm14.0$ \\
 & & & & & &\\
 $\u0$	\dotfill	& 1.960	& $\pm0.022$ & 0.6593 & $^{+0.8530}_{-0.2460}$ & 0.4462 & $^{+0.1324}_{-0.0927}$ \\
 & & & & & &\\
$\I0$  \dotfill	& 15.340 & $\pm0.003$ 	& 15.340 & $\pm0.003$ 	& 15.340 & $\pm0.003$ \\
 & & & & & &\\
 $\fsi$ \dotfill	&  1.0	& --- 			& 0.088 & $\pm0.074$	& 0.0463 & $\pm0.0154$ \\
 & & & & & &\\
 $V_0$  \dotfill	& --- 		& --- 		& --- 		& --- & 16.099 & $\pm0.008$ \\
 & & & & & &\\
 $f_{\rm S_V}$ \dotfill & --- & ---		&  ---		& --- & 0.0609 & $\pm0.0200$ \\
 & & & & & &\\
 $\chi^2$\dotfill  & 630.6 &  			 &   629.3 &   & 1610.6 &  \\
 & & & & & &\\
 ${\chi^2\over N_{dof}}$\dotfill  & 2.12	& 	&  2.13	&  & 1.47 &  \\
\hline
\end{tabular}
\end{table*}
%Nobs (I) = 301
%Nobs (I+Vtr)=1099
%%%%%%%%%%%%%%%%%%%%%%%%%%%%%%%%%%%%%%%%%%%%%%%%%%

\begin{figure}
\includegraphics[width=8.5cm]{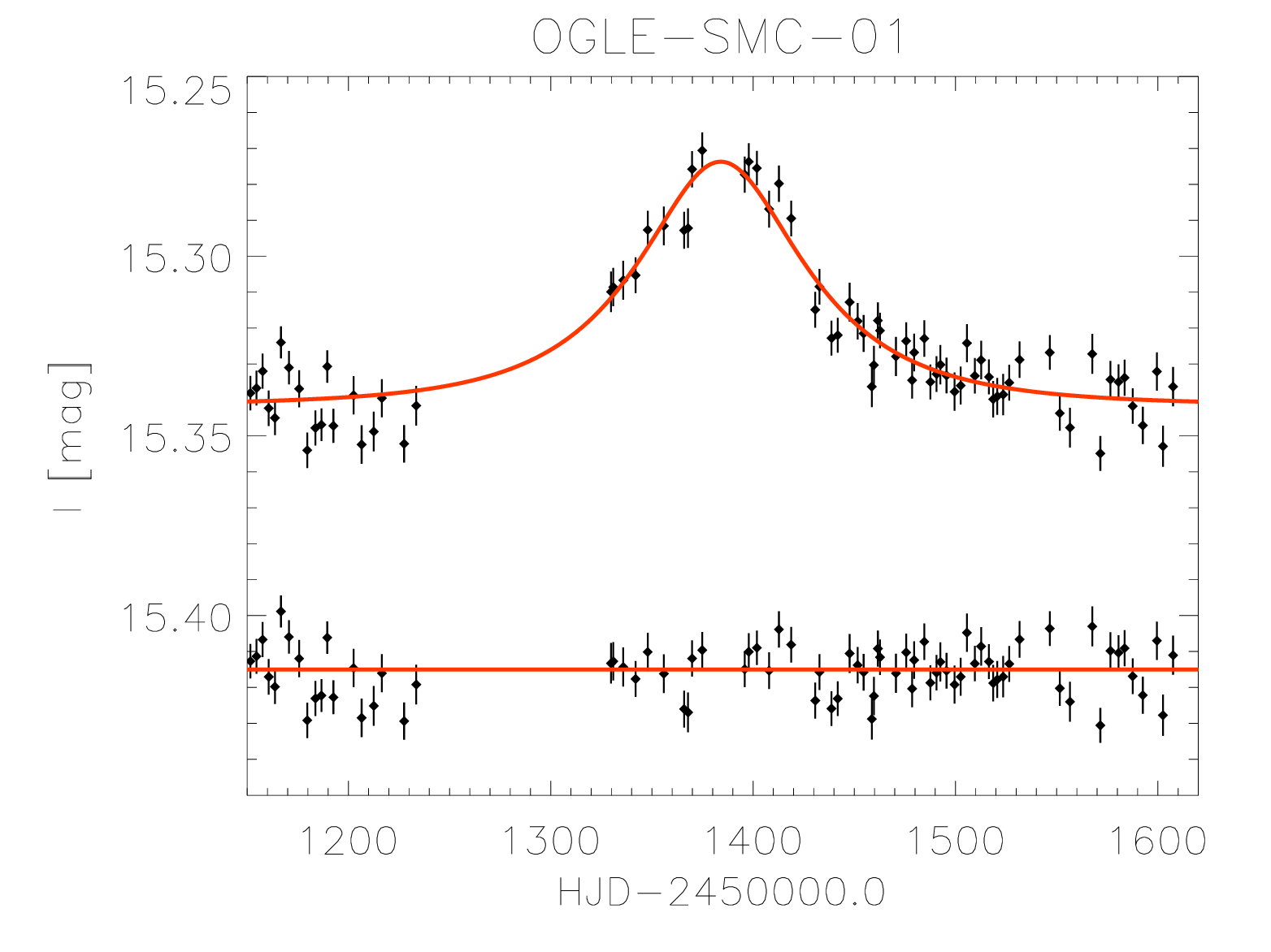}\\
\includegraphics[width=8.5cm]{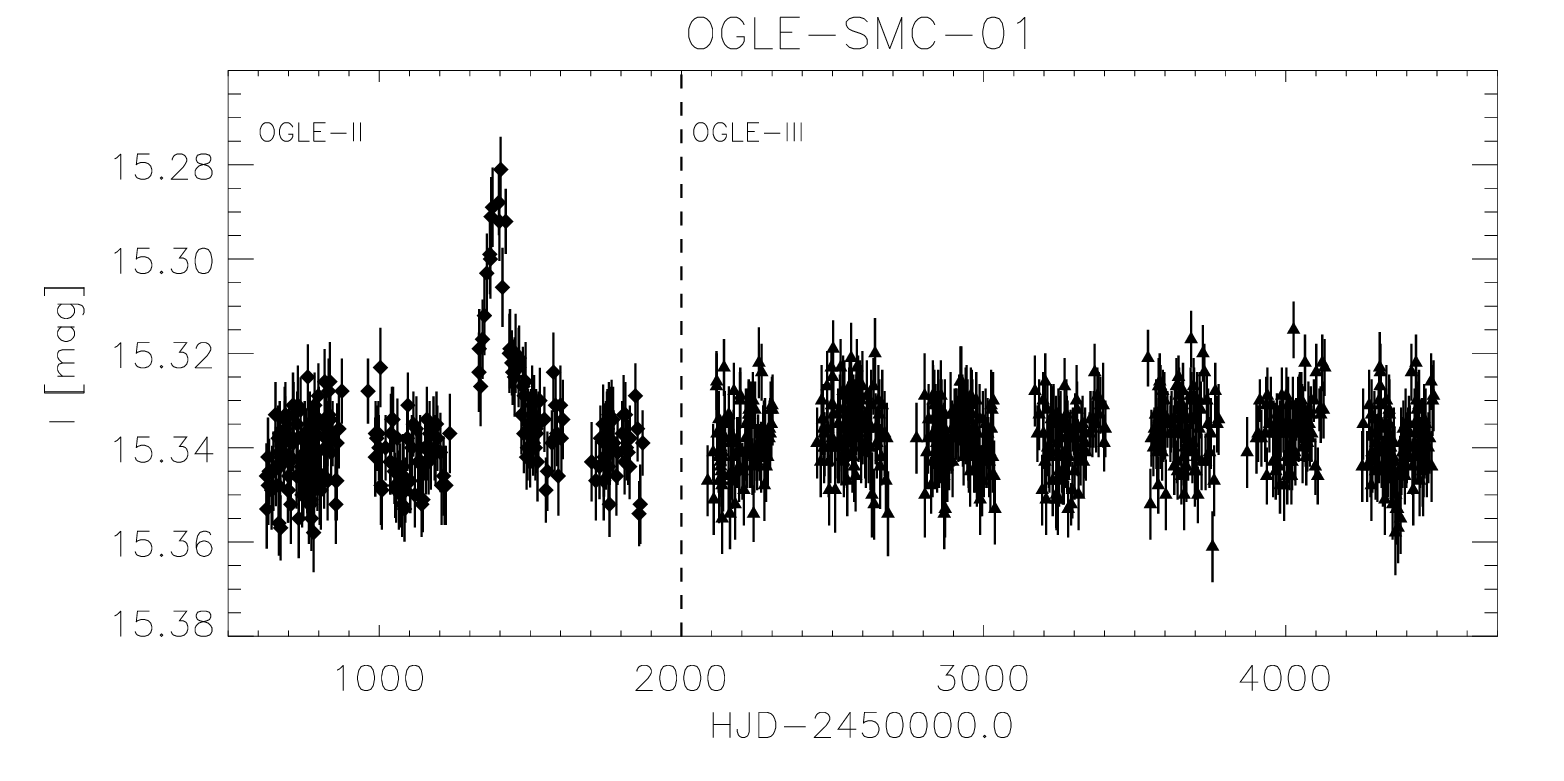}\\
\includegraphics[width=8.5cm]{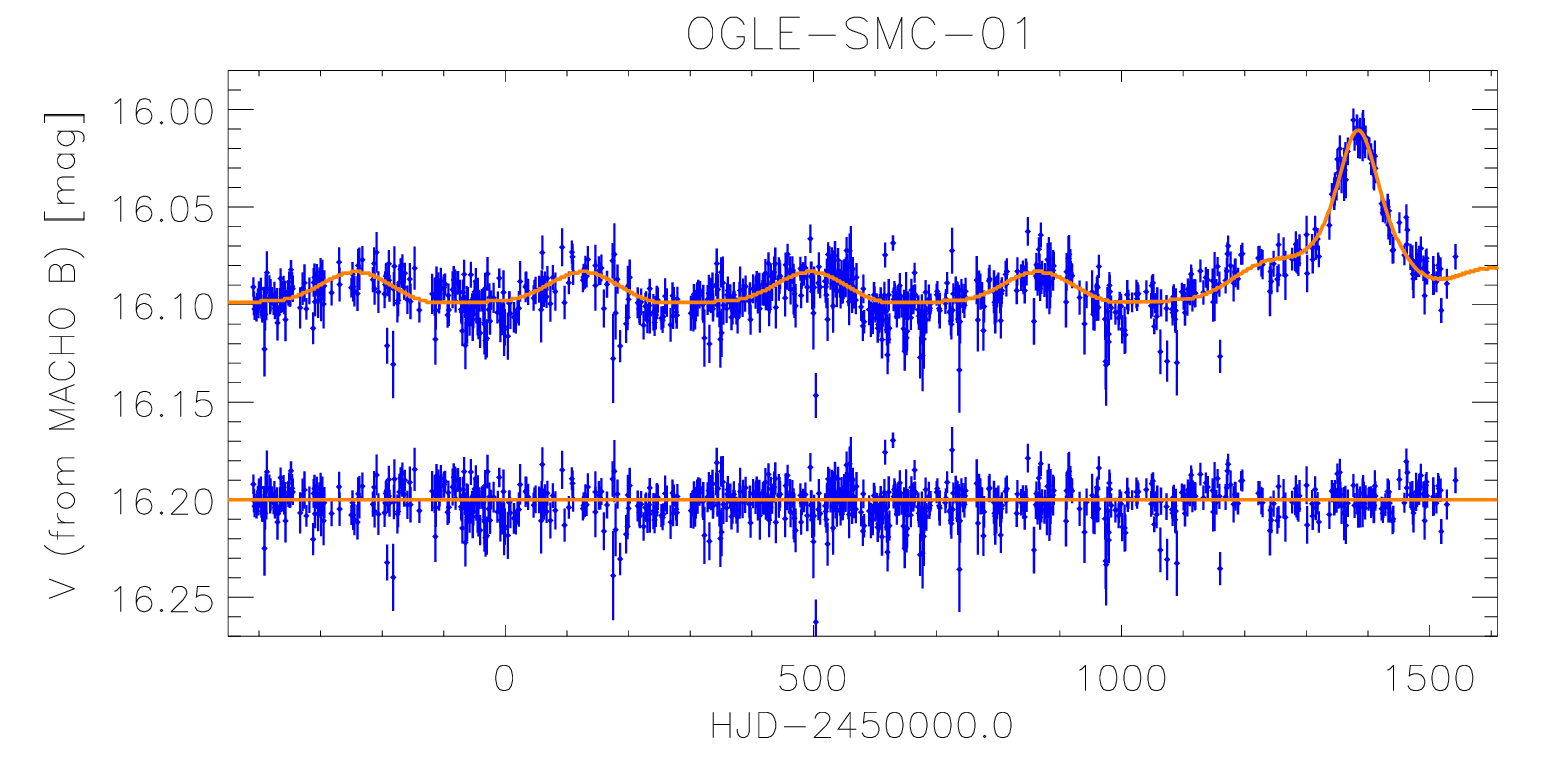}
\caption{Light curves of the OGLE-SMC-01 microlensing event candidate. {\it Top:} OGLE--II at the peak with the best model fit (seven-parameter) and its residuals. {\it Middle:} full span of OGLE--II and available OGLE--III data for that star showing no other ''bumps'' over 11 years. {\it Bottom:} full span MACHO-$B$ data transformed to the OGLE-$V$ along with the seven-parameter microlensing model and its residuals. Period of variations is close to 1 year and is not present in OGLE and EROS data.}
\label{fig:event}
\end{figure}

\begin{figure*}
\center
\includegraphics[width=8.5cm]{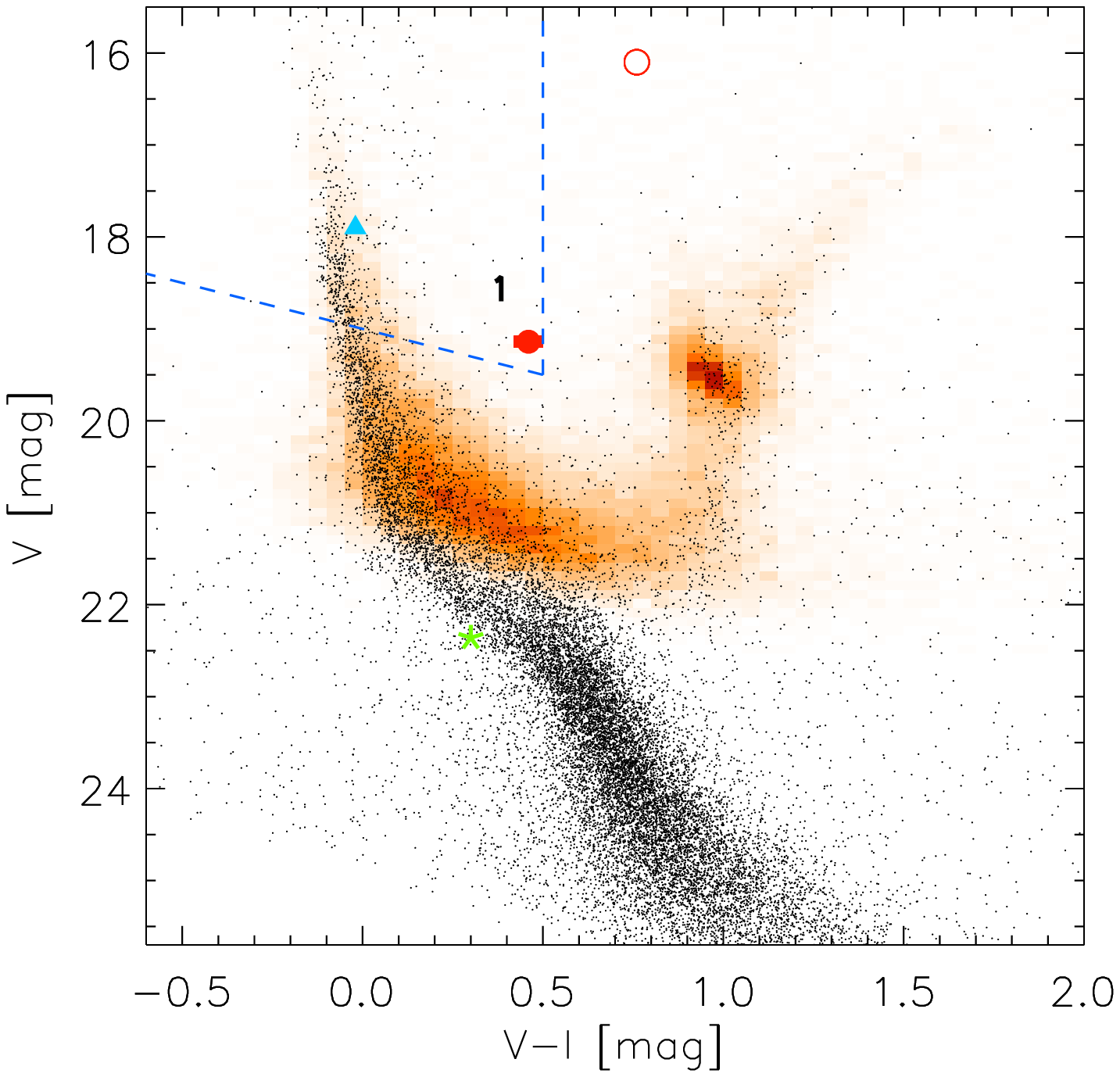}
\includegraphics[width=8.5cm]{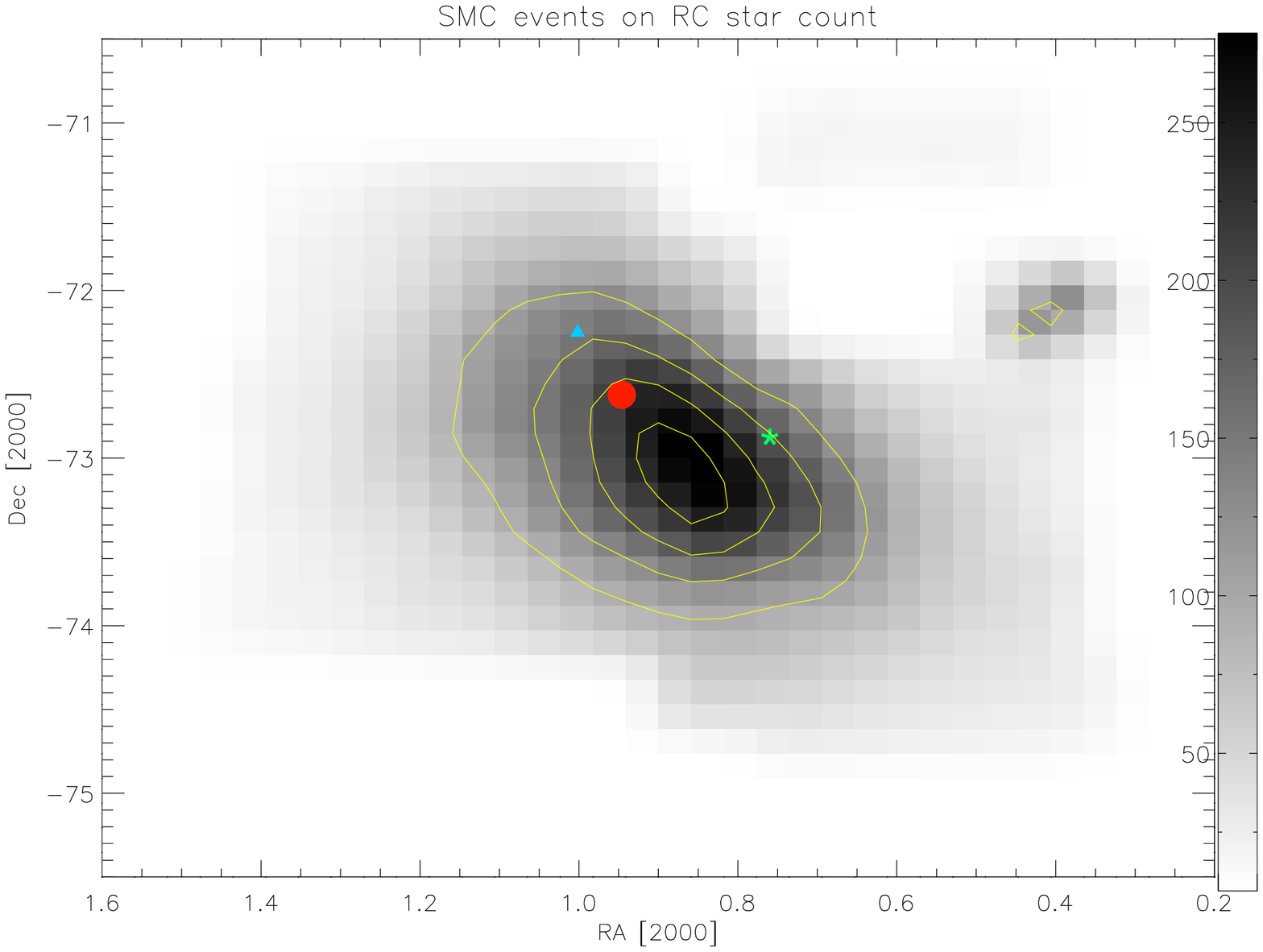}
\caption{{\it Left:} CMD of the SMC region around the OGLE--II candidate event with OGLE (red background) and {\it HST} (black dots) measurements.
The open red circle marks the baseline position of the OGLE-SMC-01 event, whereas the full dot indicates the position of the lensed source derived from the microlensing model of the event.
Events detected by MACHO are also shown: MACHO-97-SMC-1 (blue triangle) and MACHO-98-SMC-1 (green star).
Dashed lines delimit the ``Blue Bumper'' exclusion region used in our search pipeline.
{\it Right:} Positions of OGLE and MACHO candidate events on the OGLE--III SMC Red Clump stars count map with grey levels corresponding to counts per square arcminute. Yellow contours follow Red Clump density from 100 to 250 stars per sq.arcmin in intervals of 50 stars per sq.arcmin. 
}
\label{fig:cmd}
\end{figure*}

\subsection{OGLE-SMC-01}
\label{sec:event}

The OGLE-SMC-01 microlensing event $({\rm RA},{\rm Dec})=($0:56:45.89, $-$72:37:19.8) appeared on a bright blend of at least two stars with the total $I=15.340\pm0.003$ and $(V-I)=0.759\pm0.009$ in the SMC\_SC7 field. 
The $I$-band light curve of the event, measured at the difference flux position, is shown in Fig. \ref{fig:event}.

\noindent {\bf{\em MACHO data of the event}}.\\*
To interpret the nature of this event we also used the SMC data collected by the MACHO group, which are publically available\footnote{http://wwwmacho.mcmaster.ca/Data/MachoData.html}. 
In the MACHO database we found a single object (ID 207.16370.17) with the photometry available in MACHO-$B$ and MACHO-$R$ bands. 
Because of somewhat larger seeing disk in the MACHO data, as compared to OGLE, only one object was visible in the MACHO images, 
corresponding to a blend including both sources detected by OGLE. 
The original MACHO light curves showed a high level of noise and variability over entire light curve in both bands. 
To increase the signal-to-noise ratio in these light curves we re-reduced the original MACHO images using DIA in a similar manner as in \citet{Kozlowski2007}.
The red images suffered from a number of bad columns in the vicinity of our object, making it impossible for DIA to produce a reasonable light curve. 
We were able to perform the photometry in the blue channel (Fig.~\ref{fig:event}, bottom panel). As expected the DIA reduction lowered the amount of noise, as compared to the original MACHO reduction, clearly revealing a periodic variability in the baseline. With the period of 369 days (close to 1 year) this could be  some kind of artefact, such as differential refraction. The nearby stars with similar colour and brightness, however, showed no variability of that kind. 
On the other hand, the lack of any variability in the OGLE--III $V$-band data and in the EROS $B$-band data (P. Tisserand, private communication) leave us at a loss for explanation.
In the modelling we treated the variability as if it was coming from the blended star, and included it in the microlensing fit as described in \cite{varbaseline}.

In order to obtain the location of the source on a CMD, we transformed the instrumental MACHO $B$-band photometry to the standard $V$-band.
It was done by comparing the photometry of the field stars from OGLE $I$- and $V$-band, and the MACHO $B$-band. We found the following colour transformation: $(B-I) = (1.14694\pm 0.00277)(V-I)+const$, where $const$ is an artefact from instrumental magnitudes derived for MACHO data. Using this formula we were able to transform the MACHO $B$-band light curve to the standard $V$-band.
Fig. \ref{fig:event} shows the data for the event in OGLE $I$-band and $V$-band. 

\noindent {\bf{\em Microlensing model fit}}.\\*
To find the best set of parameters describing the event, we fitted a microlensing model to the available light curves in a number of ways (Table \ref{tab:fits}). 
A simple microlensing event can be successfully described with five parameters 
(Eqns 2 and 3). Our first fit was done to the OGLE $I$-band data solely, with the blending parameter $\fs$ fixed to 1, meaning that we assume that all the light comes from the microlensed source only and there are no other sources of light contributing to the overall flux.
Such a four-parameter fit almost always converges and returns good first guess parameters for the five-parameter fit. Next, we free the blending parameter and fit the light curve to obtain almost identical
goodness-of-fit values as those returned by the four-parameter fit. The derived blending indicates that only 9 per cent of the light comes from the source. Such a drop from 100 to 9 per cent for the blending parameter, and basically identical goodness of the fit, gives an overall impression of the low quality of the light curve. 

To improve the situation we performed a simultaneous fit to the OGLE $I$-band and MACHO $B$-band data.
Since microlensing is achromatic, the parameters describing the geometry of the event ($\t0,\tE,\u0$) stay the same for each data set. 
The model has only two additional parameters for an additional set of data: the blending parameter and baseline magnitude. In this model we also included the variability component to accommodate the variations present in the MACHO data, approximating their shape with 2 harmonics with period of $P=369\pm2$ days and $T_{min}=269.56\pm0.01$ for the epoch of the minimum.
Such a seven-parameter fit gives the best constraints for the parameters. The $\chi^2$ of this fit is the smallest from all three fits.

From the seven-parameter fit we obtained the colour and brightness of the source, and their formal errors as follows: 
$(V-I)_S=0.46\pm0.03$ mag and $I_S= 18.67\pm0.01$ mag. 
Note that the error-bar on the magnitude of the source is much smaller than calculated from the values and error-bars for the fitted parameters. This can be achieved because blending parameters of the fit are highly correlated.
%However, the formal errors do not include any uncertainties with respect to the possible degeneracies caused by blending. 
The microlensing model fit to the transformed $V$-band data, including the baseline's variability, is also shown in Fig. \ref{fig:event}. 
We should emphasise the fact, that the model with variability assumes the shape of the baseline variability does not change in time and thus can be predicted for the duration of the event. 
The fit to the baseline (Fig. \ref{fig:event}) seems reasonable. 
However, noise in the data can hide some irregularities of the periodicity. 
If the above assumption is not valid, namely, that the baseline variability is invariant throughout the event, there is no other way of removing the variability, and thus the colour of the source can not be derived correctly.

\noindent {\bf{\em CMD location of the source}}.\\*
Fig. \ref{fig:cmd} shows the derived location of the source (filled red dot) and the blend (open red dot) on the colour-magnitude diagram, along with two other SMC microlensing events MACHO-97-SMC-1 and MACHO-98-SMC-1. The position of OGLE-SMC-01 indicates it lies on the edge of the region we have excluded in our microlensing candidate search pipeline due to possible contamination of ``Blue Bumpers''. 
It means, if it was indeed a microlensing event and was not blended with any other star, it would have been rejected by our search pipeline. 
On the other hand, the colour and magnitude of the source were derived because the microlensing fit indicated that severe blending was present in both bands. 
If this event is not due to microlensing, we do not have any other means to de-blend the observed fluxes, thus the real CMD position of the source star in that scenario remains unknown. High-resolution imaging might help in disentangling the true nature of the source star, however, proximity of 15.3 magnitude star could be a serious obstacle.

The accumulation of problems related with this event (blending with a very bright star, artefactual variability in the MACHO data, etc.), may cast some doubts on the reliability of the colour and brightness derived for the source star. 
Also, the derived colour relies strongly on the accuracy of the model of the variability and its predictability during the event.
Therefore, with the available information, we are unable to conclude firmly whether the event is caused by microlensing and where its source star lies.

\noindent {\bf{\em Location of the blend}}.\\*
Another interesting thing to notice is the location of the blend object, as it lies in the not very populated region of the CMD. 
A plausible explanation for this could be that the bright star does not belong to the SMC and is located in the foreground, \eg a main-sequence star at a distance between 1 and 6 kpc.
Another reason for this somewhat unusual location could be that apart from the lensed source, which is bluer, there is another blue component present in the blend and the bright star itself is an SMC red giant.

\subsection{Notes on SMC events discovered by MACHO and EROS groups}
%MACHO-97-SMC-1 and MACHO-98-SMC-1 events
During OGLE--II observations of the SMC two other events happened, which were detected by the MACHO group and also reported by the EROS collaboration.
The peak of the event MACHO-97-SMC-1 \citep{Alcock1997MACHO97SMC1} occurred well before the OGLE--II observations started, therefore  
only a portion of the fading part of the light curve is present in the OGLE--II data (star SMC\_SC8.207700). 
It was not enough to be detected as a full event by our search pipeline (mainly due to Cut 6 on $\t0$, see Table \ref{tab:conditions}). 
We note in the remaining OGLE--II data, as well as in the entire 8 years of the OGLE--III data for this star, there is no other ``outburst'' visible. 
Lack of secondary deviations over many years and the presence of parallax effects in its light curve \citep{Assef2006MACHO97SMC1} confirms this is a genuine microlensing event, despite it being located in the ``Blue Bumper'' region on the CMD (see Fig. \ref{fig:cmd}).

The next event detected a year later, MACHO-98-SMC-1, aka EROS2-SMC-1 (\citealt{Alcock1999MACHO98SMC1}, \citealt{EROSMACHO98SMC1}), was an obvious caustic-crossing binary lens event, which was observed by many groups and analysed thoroughly (\eg \citealt{Afonso2000MACHO98SMC1}).
However, the source star of this event was too faint to be detected on the OGLE--II template images \citep{UdalskiMACHO98SMC1}, therefore this event is not present in the database studied here. 
However, with a baseline of about 22 mag it would not contribute to our optical depth calculations due to Cut 0 on the magnitude.

%%%%%%%%%%%%%%%% DETECTION EFFICIENCY AND BLENDING %%%%%%%%%%%%%%
\section{Blending and Detection Efficiency}
\label{sec:blending}

\begin{figure}
\center
\includegraphics[width=7.5cm]{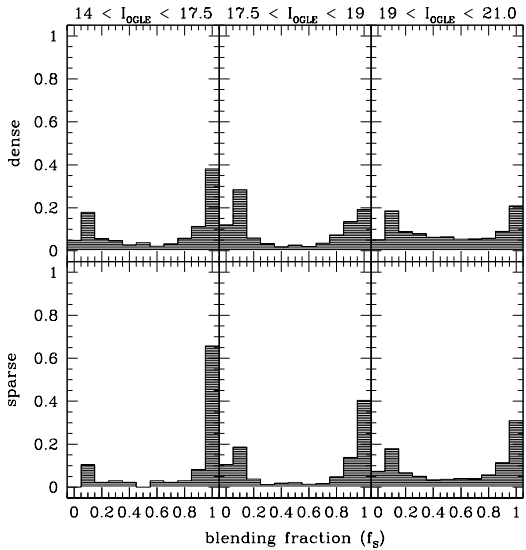}
\caption{Distributions of blending fractions for stars from the
  analysis of the archival high-resolution {\it HST} images of parts of the
  OGLE SMC fields.}
\label{fig:fs}
\end{figure}

\begin{figure}
\center
\includegraphics[width=7.5cm]{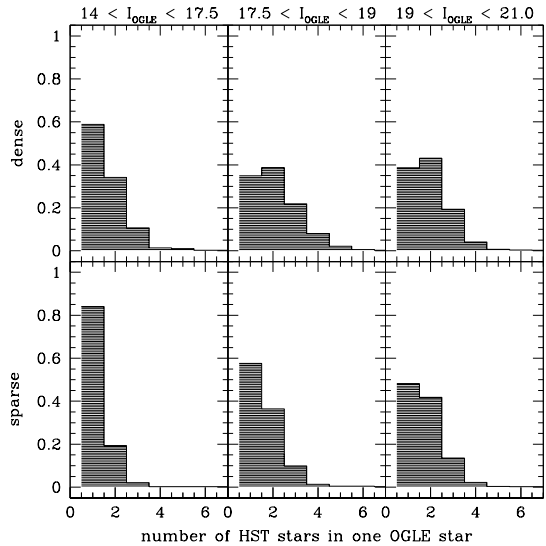}
\caption{Distribution of the number of {\it HST} stars in one OGLE--II object
  obtained for dense and sparse fields in the SMC. }
\label{fig:nhst}
\end{figure}

In this study we applied the same technique as in Paper I to deal with blending in the SMC fields. In the {\it Hubble Space Telescope} ({\it HST}) archive\footnote{http://archive.stsci.edu/hst/} we identified two images taken in the F814W filter (the closest to OGLE's $I$ band) with exposure times longer than 10 minutes, located in the dense and sparse parts of the SMC. 
Then, for each of the images we cross-matched all stars with visible objects on the corresponding OGLE image. 
This allowed us to derive mean blending distributions for dense and sparse fields and the distribution of the number of stars in each OGLE object.
Figs. \ref{fig:fs} and \ref{fig:nhst} show the distributions derived in three magnitude bins for dense and sparse OGLE fields (see Table \ref{tab:fields}).
When compared with similar distributions for the LMC, it is apparent that crowding in the SMC fields is smaller.
However, the blending is still present at some small level and affects the way we see the SMC with OGLE instruments.

Similarly, as in Paper I, we estimated the real number of monitored stars in the SMC fields using the distribution of the number of stars in a single OGLE object. 
The observed OGLE luminosity function was convolved with the blending distribution and total number of stars was calculated. 
The estimated numbers of monitored stars in each field are provided in Table \ref{tab:fields}. 
On average, the correction factor for dense fields was about 1.9 and for sparse fields about 1.7.  
The total number of monitored stars was estimated to be about 3.6 million, compared to about 2.1 million objects detected on the OGLE--II template images in Cut 0.

Derived blending distributions were then used in deriving a detection efficiency for microlensing events.
Details of the simulations and the method are provided in Paper I. 
The efficiency of event detection was obtained for each SMC field separately for time scales ranging from 1 to 1000 days. 
Because our search algorithm was not sensitive to binary lens events and other exotic events, the efficiencies were additionally corrected by a factor of 0.9 to compensate for an estimated fraction of around 10 per cent of all events being non-single.

\begin{figure}
\begin{center}
\includegraphics[width=8.5cm]{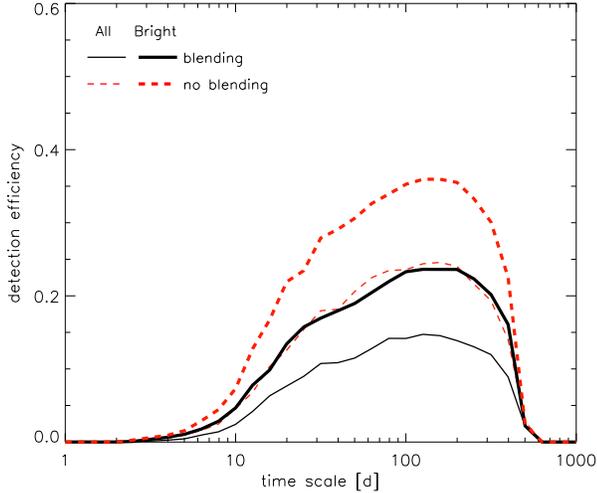}
\caption{Detection efficiencies for OGLE--II SMC\_SC7 field for All
  (thin lines) and  Bright (thick lines) Stars Samples, with blending included
  (solid lines) and neglected (dashed lines). 
}
\label{fig:eff}
\end{center}
\end{figure}

Apart from the All Stars Sample (with the magnitude cut as in the Cut 0.), the efficiencies were also derived for the Bright Sample of stars, defined as in Paper I, \ie 1 mag below the centre  of the Red Clump (usually around 19.5 mag).
Fig. \ref{fig:eff} shows an efficiency curve for the sparse field SMC\_SC7 (the one with OGLE-SMC-01) for All and Bright Stars samples, with and without blending taken into account. 
Efficiencies for the remaining fields varied within around 25 per cent from these curves, depending on the density level and the time span of observations.
As expected, the detection efficiency for bright sources is significantly higher than that for all sources. 
%%%%%%

There is also a strong dependence on taking blending into account.
If blending is neglected, the efficiency rises by approximatively 60 per cent, but the number of monitored stars is then the same as the number of objects visible on the template.
However, the smaller number of monitored stars (2.1 million compared to 3.6 million after correcting for blending, \ie 70 per cent drop) does not fully compensate the rise in the detection efficiency.
This agrees with our conclusions from Paper I that blending should be scrupulously taken into account when measuring the microlensing optical depth.

\begin{figure}
\begin{center}
\includegraphics[width=8.5cm]{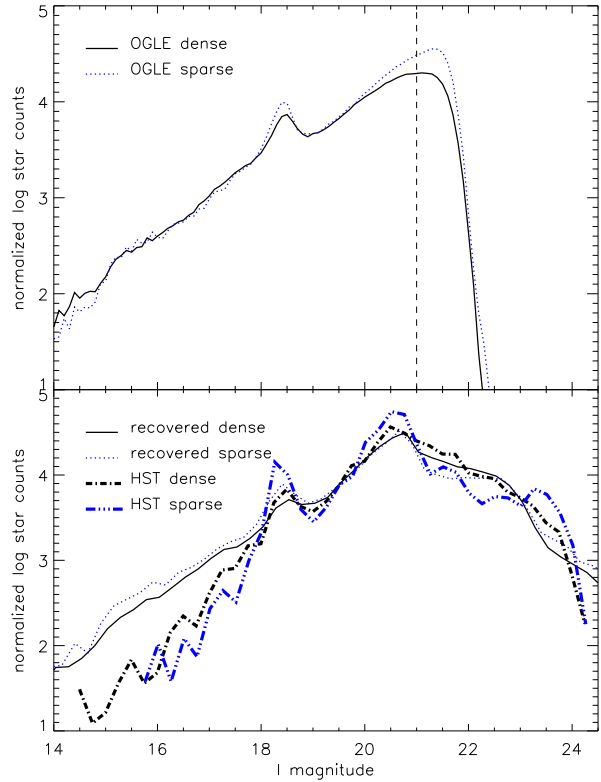}
\caption{{\it Top:} observed luminosity functions in two
  OGLE--II SMC fields, where the dotted and solid lines represent
  sparse (SC9) and dense (SC6) fields, respectively. The vertical
  line shows the cut-off at $I=21$ mag.
  {\it Bottom:} luminosity functions for the same fields recovered
  after applying the blending correction (see Section \ref{sec:blending}).
  Their shapes follow the prototype LFs of two {\it HST} fields (dense and
  sparse) used for the blending correction (thick dash-dotted lines).  }
\label{fig:lfs}
\end{center}
\end{figure}

Fig. \ref{fig:lfs} presents the luminosity functions of dense and sparse OGLE--II SMC fields (SC6 and SC9, respectively) for which blending was derived using {\it HST} archival images.
Upper panel shows functions as observed by OGLE--II, while the lower shows estimated underlying luminosity functions recovered using blending distributions from the {\it HST}. 
Blending is somewhat larger in the denser field, however, the difference here is much smaller than in the LMC (see Paper I).

%%%%%%%%%%%%%%%%% OPTICAL DEPTH %%%%%%%%%%%%%%%
\section{Optical depth estimate}
\label{sec:tau}

Even though the nature of the only candidate event we detected is unclear, 
we attempted to determine the optical depth of the single event with respect to the All Stars Sample.
In the Equation \ref{eq:tau} we used 
$T_{\rm obs}=1408$ days for the time-span of all observations,
$N_*=3.65\times10^6$ for the estimated total number of monitored stars (see Section
\ref{sec:blending}) and $N_{\rm ev}=1$ for the total number of events.

As for the time-scale of the event, we first used the value obtained in the fit to the OGLE--II $I$-band data solely 
($\te=65.0\pm21.8$) and used the efficiency  corresponding to that time-scale.
Table \ref{tab:tau} gathers all the values used in the calculations.
The error in $\tau$ was calculated with the formula given by \citet{HanGouldtau}.

For the original detection efficiency (\ie not corrected for binary events), the optical depth was derived to be
$\tau_{\rm SMC-O2} = (1.40\pm1.40)\times10^{-7}$.  
If the efficiency is corrected for non--detectability of binary lenses, the optical depth becomes $\tau_{\rm SMC-O2} = (1.55\pm1.55)\times10^{-7}$.

We also calculated the optical depth for the time-scale derived for the combination of the OGLE $I$ and MACHO $B$ data, which was slightly larger than for the OGLE data only, $\te=89.7^{+14.9}_{-13.8}$, but still within the $1\sigma$ range of the former.
Because the events' search pipeline and the detection efficiency determination did not use MACHO data, we could only estimate efficiency, assuming it was similar to the one for the OGLE $I$ data solely.
After correcting the efficiency for missed binary lenses, the optical depth was $\tau_{\rm SMC-O2} = (1.93\pm1.93)\times10^{-7}$.

\begin{table}
\caption{The optical depth for the SMC OGLE--II candidate event for time-scales derived for different data sets.}
\label{tab:tau}
\begin{center}
\begin{tabular}[h]{cccc}
\hline
event & $\te$ & $\epsilon(\te)$ &$\tau_{\rm SMC-O2}$ \\
           & [days] &                            & $\times 10^{-7}$ \\
\hline
\multicolumn{4}{c}{efficiency not corrected for binary events} \\
 & & & \\
OGLE-SMC-01 & $65.0\pm21.8$               & 0.142281 & $1.40\pm1.40$ \\
{\scriptsize (OGLE $I$ data solely)} & & & \\
\hline
OGLE-SMC-01 & $89.7^{+14.9}_{-13.8}$ & 0.157611 & $1.74\pm1.74$ \\
{\scriptsize (OGLE $I$ + MACHO $B$)} & & & \\
\hline
\hline
\multicolumn{4}{c}{efficiency corrected for binary events} \\
 & & & \\
OGLE-SMC-01 & $65.0\pm21.8$               & 0.128053 & $1.55\pm1.55$ \\
{\scriptsize (OGLE $I$ data solely)} & & & \\
\hline
OGLE-SMC-01 & $89.7^{+14.9}_{-13.8}$ & 0.141850 & $1.93\pm1.93$ \\
{\scriptsize (OGLE $I$ + MACHO $B$)} & & & \\
\hline
\end{tabular}
\end{center}
\end{table}

%%%%%%%%%%%%%%% DISCUSSION %%%%%%%%%%%%%
\section{Discussion}
\label{sec:discussion}

\subsection{On the nature of the event}

The search for microlensing events in the OGLE--II data towards the SMC yielded a discovery of a single, rather weak candidate event. 
The nature of the event is not obvious due to numerous complications in the available data (see Section \ref{sec:event}).

\noindent {\bf{\em Variable star scenario}}.\\*
The first scenario is that the event is some kind of a variable star with an occasional outburst. 
Astrometry of the observed additional flux (Fig. \ref{fig:astrometry}) indicates that the source of the outburst was severely blended with surrounding stars and 
was located in the wings of the bright star.
Therefore it was impossible to derive its real magnitude and colour and, by placing it on the CMD, to link the event to any of known outburst-like variables.
Judging just from the symmetric shape of the light curve it could be a ``Blue Bumper'' as these are known to resemble microlensing events \citep{Alcock1997}.
On the other hand, the bumpers tend to repeat their outbursts after several years (\eg \citealt{TisserandEROSLMC}), which we do not see in the available data.
The baseline of the event observed by OGLE--II and OGLE--III remained constant for over 11 years. 
The periodic variability present in the MACHO data was not observed in any other data sets and its period close to 1 year indicates it must be some sort of observational artefact.
Apart from that, MACHO data do not show any other additional outbursts, so when combined with OGLE data the event remains singular for more than 15 years of continuous observations.

\noindent {\bf{\em Microlensing scenario}}.\\*
The microlensing scenario has also its weaknesses.
Because of the severe blending with a bright star, the deviation in the light curve is rather small ($\sim~0.08$ mag).
Nevertheless, fitting to both OGLE and MACHO data (seven-parameter microlensing model) resulted in the source's brightness and colour being derived and the source's location placed on the CMD (Fig. \ref{fig:cmd}). 

The fact that the source lies in a sparsely populated region of the diagram between the main sequence branch and red clump suggests the source could be rather nearby, located in the foreground halo or the disk of our Galaxy.
However, lensing in such case is extremely unlikely, as its optical depth is of order of $10^{-9}$ (\eg \citealt{Han2008Nearfield}), \ie 2 orders of magnitude smaller than that for self-lensing of the SMC.
On the other hand, the source could be a binary in the SMC, with components of different colours blended and close enough for the microlensing not to resolve them, as no binary-source features are present in the light-curve.

Moreover, if the lensed source was located at the far end of the SMC it would suffer from high internal extinction, thus its position on the CMD would move from the main sequence towards the less populated region. 
The central location of the event on the SMC map coincides with the region where SMC is the thickest.
However, the internal extinction derived by \citet{Subramanian2009} for the region of the OGLE-SMC-01 was
$E(V-I)\approx0.08$ mag, far too small to explain the unusual position of the source.
On the other hand, the authors notice that their estimate for the extinction could be underestimated 
as the extinction in the case of the LMC, derived using main sequence OB stars as tracers \citep{Harris1997}, differs by about 0.2 mag. 
This could indicate some significant differences in the spatial distributions of different tracers within the Magellanic Clouds. 
Given the source, if reddened, belongs to the O, B or A main sequence stars, the internal extinction could reach up to $\sim 0.3$ mag.
That would be consistent with the source being located in the far end of the Cloud and self-lensed by an object from within the SMC.
Self-lensing events were already suggested to suffer from higher extinction in \citep{Zhao1999Reddening}.
%%%%%%%

Interestingly,  both other known SMC events from the period of the OGLE--II, namely MACHO-97-SMC-1 and MACHO-98-SMC-1, were considered as caused by self-lensing (\citealt{Sahu1998}, \citealt{Afonso2000MACHO98SMC1},\citealt{Assef2006MACHO97SMC1}).

The self-lensing optical depth of the SMC was estimated by \citet{Graff1999} in the N-body simulations of the SMC events.
They concluded the self-lensing contribution to the overall $\tau$ over the EROS-2 fields was $\tau_{\rm SL}\approx 0.4\times10^{-7}$, with a maximum value at the very centre of the SMC of $\tau = 1.6\times10^{-7}$. 
These estimates, however, were based on rather small line-of-sight depth of the SMC (about 2kpc) and are probably underestimated in view of recent results from \citet{Subramanian2009}, who derived the depth of the SMC bar to about 5kpc. 
Moreover, because OGLE--II covers only the very central parts of the SMC, it is more prone to see self-lensing events than EROS, which covered a much larger area. 
The long time-scale of the OGLE-SMC-01 event is also in agreement with the expected time-scale for self-lensing events of about 100 days for a mean lens mass of 0.35 $\msun$.
The time-scale of the event is also similar to those of both MACHO events, suspected to be due to self-lensing. 
Therefore, we are inclined to conclude that the most likely scenario for the OGLE-SMC-01 event is the self-lensing in the SMC.

\subsection{On the optical depth towards the SMC and the upper limit on MACHOs in the Galactic halo}

Values of the optical depth derived for different time-scales of the event, obtained using either OGLE data or a combination of OGLE and MACHO data, vary in range between 1.4 and 1.9 $10^{-7}$.
They agree with estimates for self-lensing from \citet{Graff1999} for the central parts of the SMC, but their statistical significance is very difficult to judge because of just one contributing event. 
They are, however, still in agreement with EROS's calculation for the sole event MACHO-97-SMC-1, $\tau_{\rm SMC} =  (1.7 \pm 1.7) \times10^{-7}$ \citep{TisserandEROSLMC}.
Based on their two events, the MACHO collaboration estimated $\tau=(2-3) \times 10^{-7}$ \citep{Alcock1999MACHO98SMC1}, which, again, is in agreement with our result.

\begin{figure*}
\begin{center}
\includegraphics[width=12.5cm]{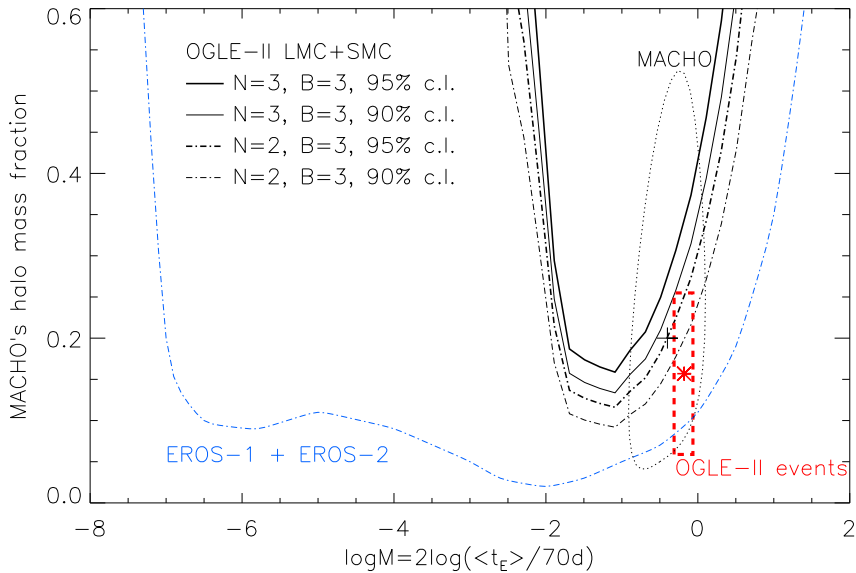}
\caption{Mass fraction in compact dark halo objects as a function of the mass of
the lensing objects for combined LMC and SMC results from OGLE--II. 
The red box with a star shows the hypothetical value for the two LMC and one SMC events if they were caused by halo dark lenses.
Solid lines show upper limits assuming all OGLE--II events are not due to MACHOs and the expected self-lensing background yields 3 events.
Black dot-dashed lines show the same upper limit, but excluding the ambiguous SMC event.
Also shown are the results of the MACHO collaboration (dotted line) and the upper limit derived by the EROS
group (dot-dashed blue line). }
\label{fig:upperlimit}
\end{center}
\end{figure*}

Yet, we must emphasise that the value of the optical depth obtained just for one single event is statistically not very significant. 
If converted to a fraction of MACHOs in the Galactic halo it would mean that between 25 and 40 per cent of halo's mass is contained in dark matter compact objects. 
Nonetheless, that would be valid only when the event we found was assumed to be caused by MACHO lensing and the modelled optical depth was $\tau_{\rm total}^S=5.1\times 10^{-7}$ for a Galaxy halo entirely constructed of MACHOs, according to model $S$ of \citet{AlcockMACHOLMC}.
Furthermore, if we speculate that all three events detected by OGLE--II towards both Magellanic Clouds were due to the MACHOs, that would mean that MACHOs compose about 15 per cent of the total mass of the Galactic halo with mean mass of 0.7 $\msun$.
This is shown as a box in Fig. \ref{fig:upperlimit}.

On the other hand, if all three events detected towards the SMC and LMC were caused by the regular luminous stellar component of each Cloud (self--lensing), there is a nil result for candidates for the MACHO lensing among 15 million monitored stars from both Clouds.
That could be translated into an upper limit for MACHO fraction in the halo using our detection efficiency and time-scale distribution for halo lenses from the model $S$.
We took the mean observing time (both Clouds were observed for the nearly the same time) and we used mean detection efficiencies to derive the total number of expected events in LMC and SMC.
The efficiency and number of monitored stars in the SMC were adjusted to be consistent with the magnitude cut applied in the LMC study (20.4 mag).
We estimated the number of events expected to be observed by OGLE--II in the SMC and LMC as caused by MACHO lenses only.
Following the suggestion of \cite{Moniez2010}, we assumed all 3 detected events are the expected self-lensing signal ($N=3$, $B=3$) and used Poisson statistics tables for such a background from \cite{FeldmanLowStatistics}.
For a ``typical'' MACHO mass of 0.4 $\msun$ we expect at least 19 events, which gives a limit of 27 per cent of the total halo mass at 95 per cent confidence.
 
However, if, based on its ambiguity, we reject the OGLE-SMC-01 as not a genuine microlensing event, there are only 2 events in both Clouds.
The exact value of the background (\ie self-lensing and other-than-MACHO lensing events) is not well known towards the SMC.
Because, however, events detected by MACHO in the 1997 and 1998 were confirmed in having a self-lensing nature, here we assume there should be at least one self-lensing event expected.
For $B=3$ and $N=2$ the upper limit on MACHO fraction in the halo becomes about 20 per cent for deflector mass of 0.4 $\msun$ and 11 per cent for masses in range 0.003 and 0.2 $\msun$, at the 95 per cent confidence level.
The upper limit on the MACHO mass fraction is shown\footnote{Data in electronic form are available through OGLE's website: {\it http://ogle.astrouw.edu.pl}} in Fig. \ref{fig:upperlimit}. 
 
\section{Conclusions}

OGLE--II has provided a new and independent constraint on the presence of compact dark matter objects in the Galactic halo. 
In the LMC there were 2 candidate events found, both most likely due to self-lensing.
A single candidate event was detected in the SMC data and its presence, if of microlensing nature at all, is consistent with the self-lensing scenario, in which the source is located at the far back end of the SMC and was lensed by a lens from within the SMC.
The unusual position of the source on the colour-magnitude diagram could also be explained by, \eg binarity of the source, and is generally not very trustworthy due to numerous ambiguities of the data.

The derived optical depth estimate for the single event indicates a value of  $\tau_{SMC}=(1.55\pm1.55)\times  10^{-7}$, which is very close to the previous measurements obtained with other data sets by the EROS and MACHO collaborations and is in agreement with self-lensing estimates.
However, its low statistical significance prevents deriving any reasonable conclusions on its origin, which may require further and more detailed studies of the structure of the SMC and resulting self-lensing optical depth.

The detection of a single and unconvincing candidate event in the OGLE--II SMC data only strengthen our previous conclusions reached in the OGLE--II LMC data (Paper I).
The hypothesis that the Galactic halo is composed of compact objects is not favoured by these results. 
They show that the fraction of compact objects is close to zero and the only reason why we cannot completely exclude the compact-object hypothesis is due to limitations of the survey.

This verdict, as determined from the OGLE--II data, will be further validated by analysing data from the recently-completed OGLE--III survey, which covers a much wider area and has a duration of 8 years.

\section*{acknowledgements}
We would like to thank for their help at various stages of this work Drs Nicholas Rattenbury, Andy Gould, Wyn Evans, Luigi Mancini, Sebastiano Calchi-Novati and Gaetano Scarpetta.
{\L}W, SK, MSC and JS acknowledge generous support from the European Community's
FR6 Marie Curie Programme, Contract No. MRTN-CT-2004-505183
``ANGLES''.  
This work was also supported by EC FR7 grant PERG04-GA-2008-234784 to {\L}W.
The OGLE project is partially supported by the Polish
MNiSW grant N20303032/4275.  JS also acknowledges support through the
Polish MNiSW grant no. N20300832/070.

\bibliographystyle{mn2e}

\begin{thebibliography}{}

\bibitem[\protect\citeauthoryear{{Afonso}, {Alard}, {Albert} \&
  {Andersen}}{{Afonso} et~al.}{2000}]{Afonso2000MACHO98SMC1}
{Afonso} C.,  {Alard} C.,  {Albert} J.~N.,    {Andersen} J. e.~a.,  2000, \apj,
  532, 340

\bibitem[\protect\citeauthoryear{{Afonso}, {Albert}, {Andersen}, {Ansari} \& et
  al.}{{Afonso} et~al.}{2003}]{EROSSMC2003}
{Afonso} C.,  {Albert} J.~N.,  {Andersen} J.,  {Ansari} R.,    et al. 2003,
  \aap, 400, 951

\bibitem[\protect\citeauthoryear{{Alard} \& {Lupton}}{{Alard} \&
  {Lupton}}{1998}]{AlardLuptonDIA}
{Alard} C.,  {Lupton} R.~H.,  1998, \apj, 503, 325

\bibitem[\protect\citeauthoryear{{Alcock}, {Akerloff}, {Allsman}, {Axelrod} \&
  et al.}{{Alcock} et~al.}{1993}]{MACHO}
{Alcock} C.,  {Akerloff} C.~W.,  {Allsman} R.~A.,  {Axelrod} T.~S.,    et al.
  1993, Nature, 365, 621

\bibitem[\protect\citeauthoryear{{Alcock}, {Allsman}, {Alves}, {Axelrod} \& et
  al.}{{Alcock} et~al.}{1997}]{Alcock1997MACHO97SMC1}
{Alcock} C.,  {Allsman} R.~A.,  {Alves} D.,  {Axelrod} T.~S.,    et al. 1997,
  \apjl, 491, L11+

\bibitem[\protect\citeauthoryear{{Alcock}, {Allsman}, {Alves} \&
  {Axelrod}}{{Alcock} et~al.}{1999}]{Alcock1999MACHO98SMC1}
{Alcock} C.,  {Allsman} R.~A.,  {Alves} D.,    {Axelrod} T.~S. e.~a.,  1999,
  \apj, 518, 44

\bibitem[\protect\citeauthoryear{{Alcock}, {Allsman}, {Alves} \& et
  al.}{{Alcock} et~al.}{1997}]{Alcock1997}
{Alcock} C.,  {Allsman} R.~A.,  {Alves} D.,    et al. 1997, \apj, 486, 697

\bibitem[\protect\citeauthoryear{{Alcock}, {Allsman}, {Alves}, {Axelrod} \& et
  al.}{{Alcock} et~al.}{2000}]{AlcockMACHOLMC}
{Alcock} C.,  {Allsman} R.~A.,  {Alves} D.~R.,  {Axelrod} T.~S.,    et al.
  2000, \apj, 542, 281

\bibitem[\protect\citeauthoryear{{Assef}, {Gould}, {Afonso}, {Albert} \& et
  al.}{{Assef} et~al.}{2006}]{Assef2006MACHO97SMC1}
{Assef} R.~J.,  {Gould} A.,  {Afonso} C.,  {Albert} J.~N.,    et al. 2006,
  \apj, 649, 954

\bibitem[\protect\citeauthoryear{{Aubourg}, {Bareyre}, {Brehin}, {Gros} \& et
  al.}{{Aubourg} et~al.}{1993}]{EROS}
{Aubourg} E.,  {Bareyre} P.,  {Brehin} S.,  {Gros} M.,    et al. 1993, Nature,
  365, 623

\bibitem[\protect\citeauthoryear{{Auri{\`e}re}, {Baillon}, {Bouquet}, {Carr} \&
  et al.}{{Auri{\`e}re} et~al.}{2001}]{POINTAGAPE}
{Auri{\`e}re} M.,  {Baillon} P.,  {Bouquet} A.,  {Carr} B.~J.,    et al. 2001,
  \apjl, 553, L137

\bibitem[\protect\citeauthoryear{{Bennett}}{{Bennett}}{2005}]{BennettMACHOLMC}
{Bennett} D.~P.,  2005, Astrophysical Journal, 633, 906

\bibitem[\protect\citeauthoryear{{Dong}, {Bond}, {Gould}, {Koz{\l}owski} \& et
  al.}{{Dong} et~al.}{2009}]{DongKB07400}
{Dong} S.,  {Bond} I.~A.,  {Gould} A.,  {Koz{\l}owski} S.,    et al. 2009,
  \apj, 698, 1826

\bibitem[\protect\citeauthoryear{{EROS Collaboration}, {Afonso}, {Alard},
  {Albert} \& et al.}{{EROS Collaboration} et~al.}{1998}]{EROSMACHO98SMC1}
{EROS Collaboration} {Afonso} C.,  {Alard} C.,  {Albert} J.~N.,    et al. 1998,
  \aap, 337, L17

\bibitem[\protect\citeauthoryear{{Feldman} \& {Cousins}}{{Feldman} \&
  {Cousins}}{1998}]{FeldmanLowStatistics}
{Feldman} G.~J.,  {Cousins} R.~D.,  1998, Physical Review D, 57, 3873

\bibitem[\protect\citeauthoryear{{Gaudi}, {Bennett}, {Udalski} \&
  {Gould}}{{Gaudi} et~al.}{2008}]{Gaudi2008}
{Gaudi} B.~S.,  {Bennett} D.~P.,  {Udalski} A.,    {Gould} A. e.~a.,  2008,
  Science, 319, 927

\bibitem[\protect\citeauthoryear{{Gould}}{{Gould}}{2000}]{Gould2000microlensingFormalism}
{Gould} A.,  2000, \apj, 542, 785

\bibitem[\protect\citeauthoryear{{Gould}, {Udalski}, {Monard}, {Horne}, {Dong},
  {Miyake}, {Sahu}, {Bennett}, {Wyrzykowski}, {Soszy{\'n}ski}, {Szyma{\'n}ski},
  {Kubiak}, {Pietrzy{\'n}ski}, {Szewczyk}, {Ulaczyk} \& et al.}{{Gould}
  et~al.}{2009}]{Gould2009terrestialParallax}
{Gould} A.,  {Udalski} A.,  {Monard} B.,  {Horne} K.,  {Dong} S.,  {Miyake} N.,
   {Sahu} K.,  {Bennett} D.~P.,  {Wyrzykowski} {\L}.,  {Soszy{\'n}ski} I.,
  {Szyma{\'n}ski} M.~K.,  {Kubiak} M.,  {Pietrzy{\'n}ski} G.,  {Szewczyk} O.,
  {Ulaczyk} K.,    et al. 2009, \apjl, 698, L147

\bibitem[\protect\citeauthoryear{{Graff} \& {Gardiner}}{{Graff} \&
  {Gardiner}}{1999}]{Graff1999}
{Graff} D.~S.,  {Gardiner} L.~T.,  1999, \mnras, 307, 577

\bibitem[\protect\citeauthoryear{{Han}}{{Han}}{2008}]{Han2008Nearfield}
{Han} C.,  2008, \apj, 681, 806

\bibitem[\protect\citeauthoryear{{Han} \& {Gould}}{{Han} \&
  {Gould}}{1995}]{HanGouldtau}
{Han} C.,  {Gould} A.,  1995, \apj, 449, 521

\bibitem[\protect\citeauthoryear{{Harris}, {Zaritsky} \& {Thompson}}{{Harris}
  et~al.}{1997}]{Harris1997}
{Harris} J.,  {Zaritsky} D.,    {Thompson} I.,  1997, \aj, 114, 1933

\bibitem[\protect\citeauthoryear{{Kerins}}{{Kerins}}{2008}]{ANGSTROM}
{Kerins} E.,  2008, The Astronomer's Telegram, 1857, 1

\bibitem[\protect\citeauthoryear{{Koz{\l}owski}, {Wo{\'z}niak}, {Mao} \&
  {Wood}}{{Koz{\l}owski} et~al.}{2007}]{Kozlowski2007}
{Koz{\l}owski} S.,  {Wo{\'z}niak} P.~R.,  {Mao} S.,    {Wood} A.,  2007, \apj,
  671, 420

\bibitem[\protect\citeauthoryear{{Mao}, {Smith}, {Wo{\'z}niak}, {Udalski},
  {Szyma{\'n}ski}, {Kubiak}, {Pietrzy{\'n}ski}, {Soszy{\'n}ski} \&
  {{\.Z}ebru{\'n}}}{{Mao} et~al.}{2002}]{OGLEBH}
{Mao} S.,  {Smith} M.~C.,  {Wo{\'z}niak} P.,  {Udalski} A.,  {Szyma{\'n}ski}
  M.,  {Kubiak} M.,  {Pietrzy{\'n}ski} G.,  {Soszy{\'n}ski} I.,
  {{\.Z}ebru{\'n}} K.,  2002, \mnras, 329, 349

\bibitem[\protect\citeauthoryear{{Moniez}}{{Moniez}}{2010}]{Moniez2010}
{Moniez} M.,  2010, preprint, (arXiv/1001.2707)

\bibitem[\protect\citeauthoryear{{Paczy{\'n}ski}}{{Paczy{\'n}ski}}{1986}]{Paczynski1986}
{Paczy{\'n}ski} B.,  1986, \apj, 304, 1

\bibitem[\protect\citeauthoryear{{Paczy{\'n}ski}}{{Paczy{\'n}ski}}{1996}]{Paczynski1996}
{Paczy{\'n}ski} B.,  1996, \araa, 34, 419

\bibitem[\protect\citeauthoryear{{Palanque-Delabrouille}, {Afonso}, {Albert} \&
  et al.}{{Palanque-Delabrouille} et~al.}{1998}]{EROSSMC1998}
{Palanque-Delabrouille} N.,  {Afonso} C.,  {Albert} J.~N.,    et al. 1998,
  \aap, 332, 1

\bibitem[\protect\citeauthoryear{{Pojmanski}}{{Pojmanski}}{1997}]{ASAS}
{Pojmanski} G.,  1997, Acta Astronomica, 47, 467

\bibitem[\protect\citeauthoryear{{Riffeser}, {Fliri}, {Bender}, {Seitz} \&
  {G{\"o}ssl}}{{Riffeser} et~al.}{2003}]{WECAPP}
{Riffeser} A.,  {Fliri} J.,  {Bender} R.,  {Seitz} S.,    {G{\"o}ssl} C.~A.,
  2003, \apjl, 599, L17

\bibitem[\protect\citeauthoryear{{Sahu} \& {Sahu}}{{Sahu} \&
  {Sahu}}{1998}]{Sahu1998}
{Sahu} K.~C.,  {Sahu} M.~S.,  1998, \apjl, 508, L147

\bibitem[\protect\citeauthoryear{{Skowron}, {Jaroszynski}, {Udalski}, {Kubiak},
  {Szyma{\'n}ski}, {Pietrzy{\'n}ski}, {Soszy{\'n}ski}, {Szewczyk},
  {Wyrzykowski} \& {Ulaczyk}}{{Skowron} et~al.}{2007}]{Skowron2007binaries}
{Skowron} J.,  {Jaroszynski} M.,  {Udalski} A.,  {Kubiak} M.,  {Szyma{\'n}ski}
  M.~K.,  {Pietrzy{\'n}ski} G.,  {Soszy{\'n}ski} I.,  {Szewczyk} O.,
  {Wyrzykowski} {\L}.,    {Ulaczyk} K.,  2007, Acta Astronomica, 57, 281

\bibitem[\protect\citeauthoryear{{Smith}, {Mao} \& {Paczy{\'n}ski}}{{Smith}
  et~al.}{2003}]{Smith2003parallax}
{Smith} M.~C.,  {Mao} S.,    {Paczy{\'n}ski} B.,  2003, \mnras, 339, 925

\bibitem[\protect\citeauthoryear{{Subramanian} \& {Subramaniam}}{{Subramanian}
  \& {Subramaniam}}{2009}]{Subramanian2009}
{Subramanian} S.,  {Subramaniam} A.,  2009, \aap, 496, 399

\bibitem[\protect\citeauthoryear{{Szyma{\'n}ski}}{{Szyma{\'n}ski}}{2005}]{Szymanski2005}
{Szyma{\'n}ski} M.~K.,  2005, Acta Astronomica, 55, 43

\bibitem[\protect\citeauthoryear{{Tisserand}, {Le Guillou}, {Afonso}, {Albert},
  {The EROS-2 Collaboration} \& {et al.}}{{Tisserand}
  et~al.}{2007}]{TisserandEROSLMC}
{Tisserand} P.,  {Le Guillou} L.,  {Afonso} C.,  {Albert} J.~N.,  {The EROS-2
  Collaboration}   {et al.} 2007, \aap, 469, 387

\bibitem[\protect\citeauthoryear{{Udalski}}{{Udalski}}{2003}]{EWSOGLE3}
{Udalski} A.,  2003, Acta Astronomica, 53, 291

\bibitem[\protect\citeauthoryear{{Udalski}, {Jaroszy{\'n}ski}, {Paczy{\'n}ski}
  \& et al.}{{Udalski} et~al.}{2005}]{UdalskiOB05071}
{Udalski} A.,  {Jaroszy{\'n}ski} M.,  {Paczy{\'n}ski} B.,    et al. 2005,
  \apjl, 628, L109

\bibitem[\protect\citeauthoryear{{Udalski}, {Kubiak} \&
  {Szyma{\'n}ski}}{{Udalski} et~al.}{1997}]{OGLE2}
{Udalski} A.,  {Kubiak} M.,    {Szyma{\'n}ski} M.,  1997, Acta Astronomica, 47,
  319

\bibitem[\protect\citeauthoryear{{Udalski}, {Szyma{\'n}ski}, {Ka{\l}u{\.z}ny},
  {Kubiak}, {Krzemi{\'n}ski}, {Mateo}, {Preston} \& {Paczy{\'n}ski}}{{Udalski}
  et~al.}{1993}]{Udalski1993}
{Udalski} A.,  {Szyma{\'n}ski} M.,  {Ka{\l}u{\.z}ny} J.,  {Kubiak} M.,
  {Krzemi{\'n}ski} W.,  {Mateo} M.,  {Preston} G.~W.,    {Paczy{\'n}ski} B.,
  1993, Acta Astronomica, 43, 289

\bibitem[\protect\citeauthoryear{{Udalski}, {Szymanski}, {Kubiak},
  {Pietrzynski}, {Wozniak} \& {Zebrun}}{{Udalski}
  et~al.}{1997}]{UdalskiMACHO98SMC1}
{Udalski} A.,  {Szymanski} M.,  {Kubiak} M.,  {Pietrzynski} G.,  {Wozniak} P.,
    {Zebrun} K.,  1997, Acta Astronomica, 47, 431

\bibitem[\protect\citeauthoryear{{Wo{\'z}niak}}{{Wo{\'z}niak}}{2000}]{WozniakDIA}
{Wo{\'z}niak} P.~R.,  2000, Acta Astronomica, 50, 421

\bibitem[\protect\citeauthoryear{{Wyrzykowski}, {Koz{\l}owski}, {Skowron},
  {Belokurov}, {Smith}, {Udalski}, {Szyma{\'n}ski}, {Kubiak},
  {Pietrzy{\'n}ski}, {Soszy{\'n}ski}, {Szewczyk} \&
  {{\.Z}ebru{\'n}}}{{Wyrzykowski} et~al.}{2009}]{Wyrzykowski2009}
{Wyrzykowski} {\L}.,  {Koz{\l}owski} S.,  {Skowron} J.,  {Belokurov} V.,
  {Smith} M.~C.,  {Udalski} A.,  {Szyma{\'n}ski} M.~K.,  {Kubiak} M.,
  {Pietrzy{\'n}ski} G.,  {Soszy{\'n}ski} I.,  {Szewczyk} O.,
  {{\.Z}ebru{\'n}} K.,  2009, \mnras, 397, 1228

\bibitem[\protect\citeauthoryear{{Wyrzykowski}, {Udalski}, {Mao}, {Kubiak},
  {Szyma{\'n}ski}, {Pietrzy{\'n}ski}, {Soszy{\'n}ski} \&
  {Szewczyk}}{{Wyrzykowski} et~al.}{2006}]{varbaseline}
{Wyrzykowski} {\L}.,  {Udalski} A.,  {Mao} S.,  {Kubiak} M.,  {Szyma{\'n}ski}
  M.~K.,  {Pietrzy{\'n}ski} G.,  {Soszy{\'n}ski} I.,    {Szewczyk} O.,  2006,
  Acta Astronomica, 56, 145

\bibitem[\protect\citeauthoryear{{Yock}}{{Yock}}{1998}]{MOA}
{Yock} P.~C.~M.,  1998, in {Sato} H.,  {Sugiyama} N.,  eds, Frontiers Science
  Series 23: Black Holes and High Energy Astrophysics. UAP, Tokyo, p.~375

\bibitem[\protect\citeauthoryear{{Zhao}}{{Zhao}}{1999}]{Zhao1999Reddening}
{Zhao} H.,  1999, \apj, 527, 167

\end{thebibliography}

\label{lastpage}
\end{document}